\documentclass[aip,avs,reprint,superscriptaddress,longbibliography,nofootinbib]{revtex4-2}

\usepackage{graphicx}  
\usepackage{amsmath}   
\usepackage{amssymb}   
\usepackage{amsfonts}  
\usepackage{bm}        
\usepackage{dcolumn}   
\usepackage{physics}   
\usepackage[utf8]{inputenc} 
\usepackage[T1]{fontenc}   
\usepackage{subcaption}

\usepackage{hyperref}
\hypersetup{
    colorlinks=true,
    linkcolor=blue,
    filecolor=blue,
    urlcolor=blue,
    citecolor=blue
}

\begin{document}

\title{Generalized State Discrimination for Tunable Quantum Key Distribution}

\author{Animesh Banik}
    \email{animesh4physics@gmail.com} 
    \thanks{Corresponding author}   
    \affiliation{QRNLab, Department of Physics, University of Chittagong, Chittagong 4331, Bangladesh}

\author{Md. Shihab Khan}
    \affiliation{QRNLab, Department of Physics, University of Chittagong, Chittagong 4331, Bangladesh}

\author{Rafid Masrur Khan}
    \affiliation{Department of Electrical and Computer Engineering, North South University, Dhaka 1229, Bangladesh}

\author{Syed Emad Uddin Shubha}
    \affiliation{Division of Computer Science, Louisiana State University, Baton Rouge, Louisiana 70803, USA}

\author{Quazi Muhammad Rashed Nizam}
    \affiliation{QRNLab, Department of Physics, University of Chittagong, Chittagong 4331, Bangladesh}

\date{\today}


\begin{abstract}
We introduce a tunable framework for generalized quantum state discrimination (GSD) and apply it to quantum key distribution (QKD) through a protocol we call \textit{phiQKD}. Building upon the two-state B92 protocol, phiQKD replaces the traditional unambiguous state discrimination (USD) measurement with a one-parameter family of hybrid POVMs characterized by a tilting angle $\phi$. This allows for continuous control over the trade-off among correct, incorrect, and inconclusive outcomes. While offering improvement in key rate over B92, the primary practical advantage of phiQKD lies in its adaptability to noise and channel imperfections via measurement tunability. By evaluating the protocol under asymptotic, finite-key, and composable security models, we show that, treating quantum measurement as a tunable design parameter, rather than a fixed operation, enables flexible protocol optimization and improved performance under realistic constraints.  
\end{abstract}

\maketitle

\section{Introduction}
In an era where quantum algorithms capable of breaking classical cryptographic routines exist and are close to being physically realizable, is any information secure? The Internet, the backbone of the digital and modern age, relies on classical public key cryptography such as RSA \cite{rivest1978method} and Elliptic Curve Cryptography (ECC) \cite{miller1985use} to account for information security. Such systems depend on the computational difficulty of mathematical problems e.g., integer factorization and the discrete logarithm problem, to guarantee security. However, the coming of quantum computing, first conceptualized by Feynman \cite{Feynman1982, feynman1986quantum}, poses a severe threat to these foundations \cite{mosca2018cybersecurity}. For example, Shor's algorithm \cite{shor1994algorithms, shor1999polynomial}, a quantum algorithm that can factor large integers and solve discrete logarithm problems as demonstrated in \cite{gidney2021factor}, is exponentially faster than the best known classical algorithms. This implies that sufficiently powerful quantum computers could break currently employed cryptographic systems with ease \cite{nielsen2010quantum}. 

If the security of information is at risk, the only way left is to look for ``post-quantum'' or ``quantum-safe'' cryptographic solutions \cite{bernstein2025post}. While one approach involves developing classical algorithms believed to be resistant to quantum attacks (Post-Quantum Cryptography), another way to traverse would be to incorporate the laws of quantum mechanics, the underlying laws of the universe, in these security routines, ensuring that these next-generation quantum computers cannot threaten the security of these systems.

This is the domain of Quantum Key Distribution (QKD) \cite{gisin2002quantum, scarani2009security}. QKD allows two parties, traditionally named Alice and Bob, to establish a shared, secret random key. The security of such protocols relies on the probabilistic nature of quantum measurements. More specifically, while orthogonal quantum states can be distinguished perfectly using projective measurements, practical quantum protocols often involve non-orthogonal states, which resist exact identification due to the linearity of quantum mechanics and the no-cloning theorem \cite{Wootters1982}. Any attempt made by an eavesdropper, traditionally named Eve, will result in detectable changes in the system and alert the legitimate parties about the presence of an intruder, thus ensuring security. Hence, the task of identifying the correct state from a known set of states, the quantum state discrimination problem, has become foundational in quantum information, central to tasks such as quantum communication, key distribution, and teleportation \cite{chefles2000quantum,bergou200411}.

To overcome this problem, several discrimination strategies have been developed. The minimum-error strategy proposed by Helstrom \cite{helstrom1969quantum} minimizes the average error probability by projecting onto an optimal basis. Unambiguous State Discrimination (USD), introduced by Ivanovic, Dieks, and Peres \cite{ivanovic1987differentiate,dieks1988overlap,peres1988differentiate}, avoids errors altogether by allowing for inconclusive results, implemented via Positive Operator-Valued Measures (POVMs). These two strategies represent extremes in a broader trade-off landscape of correctness versus conclusiveness.

The FRIO (Fixed Rate of Inconclusive Outcome) approach extends this landscape by minimizing the error probability under a fixed inconclusive rate constraint \cite{herzog2012optimal,bagan2012optimal}. It allows for tunable trade-offs between detection efficiency and accuracy, though it still requires a pre-specified constraint on inconclusive outcomes. So in this paper, we introduce a tunable framework: Generalized State Discrimination (GSD), that generates a continuous family of POVMs using a single geometric parameter: the \textit{tilting angle} $\phi$. This formulation smoothly interpolates between USD and minimum-error discrimination, without predefining constraints. Within this framework, we identify two meaningful points Confidence Threshold Point (CTP), Equal Risk Point (ERP) where the probabilities of correct, incorrect and inconclusive outcomes exhibit balance. 

To demonstrate its practical relevance, we embed the GSD approach into a two-state QKD protocol derived from B92 \cite{bennett1992communication}, which we call phiQKD. The B92 protocol typically employs USD, resulting in high rates of inconclusive outcomes. Our tunable POVM-based strategy enables phiQKD to dynamically adjust this trade-off, yielding improved QBER suppression and higher sifting efficiency in realistic conditions. We then evaluate the performance of phiQKD across various tilting angles. We derive key rates under asymptotic, finite-key, and composable security models using the Devetak--Winter bound \cite{devetak2005distillation}, the entropic uncertainty relation with quantum memory \cite{berta2010uncertainty}, and finite-size security techniques \cite{renner2008security, hoeffding1963probability,serfling1974probability,tomamichel2012tight}. For a representative configuration using the signal states $|0\rangle$ and $|+\rangle$, we obtain a composable secure key rate of 0.181958 bits per signal, exceeding the theoretical secure key rate of standard B92 (0.156862), with lower QBER and improved throughput of about 16\%. This highlights the robustness and deployability of phiQKD under realistic noise and loss.

Beyond QKD, the GSD formalism opens up avenues for broader applications in quantum hypothesis testing, quantum metrology, and adaptive sensing \cite{eldar2003semidefinite,namkung2020generalized,becerra2013implementation}. It also connects to emerging studies on tunable measurement design in quantum protocols, including hybrid strategies for dynamic decision-making and change-point detection \cite{mohan2023generalized}.

This work does not introduce but rather promotes a shift in perspective: quantum measurement need not be a static operation, but can instead be treated as a tunable design element. By leveraging this viewpoint, we demonstrate a new pathway for optimizing QKD protocols, one that balances theoretical guarantees with practical flexibility.

\section{Background Studies and Technical Overview}

\subsection{Limitations of Existing Quantum Key Distribution (QKD)}
Although widely regarded QKD protocols, such as BB84 and B92, provide information-theoretic security, there remain limitations on the practical efficiency and robustness due to the measurement strategies used. In particular, the B92 protocol depends heavily on Unambiguous State Discrimination (USD). In the ideal case, USD is capable of accurate discrimination between states by allowing a certain probability of inconclusive detections (where no information about the state to be detected is gained by the receiver). But this works at great cost:
\begin{itemize}
    \item High Inconclusive Rates: States with significant overlap, i.e., non-orthogonal states that are close in proximity, pose the problem of a significant portion of inconclusive outcomes in USD method. In real scenarios, this results in a huge amount of discarding of qubits and hence a low rate of key generation.
    \item Susceptibility to Loss and Noise: During experimental realization of such schemes, noise is a paramount concern for the quantum channel. For channels with photon loss and dark counts, the requirement of absolute accuracy becomes a disadvantage. In such cases, the protocol decides to discard any and all counts that could lead to error, thus resulting in a tremendous percentage of signals being lost in the process.
    \item Fixed Measurement Basis: Most protocols, particularly the traditional ones, employ fixed measurement techniques (e.g., projective measurements on specific bases or fixed POVMs). This restricts any flexibility to adjust to fluctuating channel conditions. For example, it might be beneficial to tolerate a slightly higher error rate (which can be corrected during post-processing) in exchange for a higher detection rate. On the other hand, in a low-noise environment, we might prioritize lower error rates. Very few existing protocols offer such tunability.
\end{itemize} 
Although the MED strategy allows maximum conclusive detection, it also comes with a large probability of erroneous results that might negatively impact the key generation rate even after post-processing. It is particularly detrimental to security proofs in QKD because keeping the Quantum Bit Error Rate as low as possible is standard practice. Currently, there is a lack of a unified framework that allows for a continuous, tunable trade-off between these extremes within a QKD protocol, enabling it to adapt dynamically to the operational environment.

\subsection{Quantum State Discrimination}
The problem is defined as detecting the correct state among a set of quantum states that is sent from one party to another. Quantum state discrimination protocols are categorized based on the trade-off they make between the probability of error and the allowance of inconclusive outcomes. These protocols are essential in applications involving non-orthogonal quantum states, where perfect discrimination is forbidden by the structure of quantum mechanics.

Let $\ket{\psi_1},\ket{\psi_2}$ be two non-orthogonal pure states with known a priori probabilities. The goal of a discrimination protocol is to design a measurement, projective or generalized (POVM), that extracts the maximum amount of information about the input state, according to specific optimality criteria. Three primary discrimination strategies are commonly studied in this regard:

\begin{itemize}
    \item Minimum-Error Discrimination (MED) (Helstrom): Optimizes measurement to minimize the overall probability of error, with no allowance for inconclusive results.

    \item Unambiguous State Discrimination (USD): Allows for an inconclusive outcome but guarantees zero probability of error in conclusive outcomes.

    \item Intermediate or Interpolating Strategies (e.g., FRIO): Aim to balance both error and inconclusive rates under some cost function.
\end{itemize}

Each of these strategies leads to different measurement bases and performance characteristics. In the following subsections, we explore the measurement frameworks: PVM and POVM, and analyze the distinctions between the Helstrom, USD, and FRIO approaches, setting the stage for our proposed Generalized State Discrimination (GSD) scheme.

\subsection{PVM, POVM(USD), Helstrom(MED) and FRIO Approach}
\subsubsection{Projective Measurement (PVM)}
A set of orthogonal projection operators $\{P_m\}$ that act on the Hilbert space of the system is used to describe these types of measurements, which are often known as Von Neumann measurements. These operators must satisfy two conditions: 
\begin{enumerate}
    \item The completeness relation $\sum_m P_m = I$ (where $I$ is the identity operator) and 
    \item The Orthogonality $P_m P_k = \delta_{mk} P_m$.
\end{enumerate} 
If $\rho$ is the density matrix describing the state of a system, the probability of achieving outcome $m$ is given by $p(m) = \operatorname{Tr}(\rho P_m)$. if outcome $m$ is obtained , the state collapses to $\rho_m = \frac{P_m \rho P_m^\dagger}{\operatorname{Tr}(\rho P_m)}$. Projective measurements are adequate for distinguishing linearly independent (orthogonal) quantum states perfectly. 
\subsubsection{Positive Operator-Valued Measure (POVM) or USD Approach}
A more general framework known as POVM is used for quantum measurements. In this case, a set of positive semi-definite operators (POVM elements) $\{\Pi_i\}$ is used to describe a POVM. These elements sum to the identity operator, $\sum_{i=0}^n \Pi_i = I$. The probability of getting outcome $i$ when measuring a state $\rho$ is $p(i) = \operatorname{Tr}(\rho \Pi_i)$. Note that, among the $(n+1)$ POVM elements $\Pi_0$ is mapped to represent the inconclusive events. Unlike projective measurements, the POVM elements $\Pi_i$ do not need to be orthogonal projectors. This generalization equips POVMs with the capability to discriminate between non-orthogonal quantum states.

\begin{figure}[t]
    \centering
    \includegraphics[width=1\linewidth]{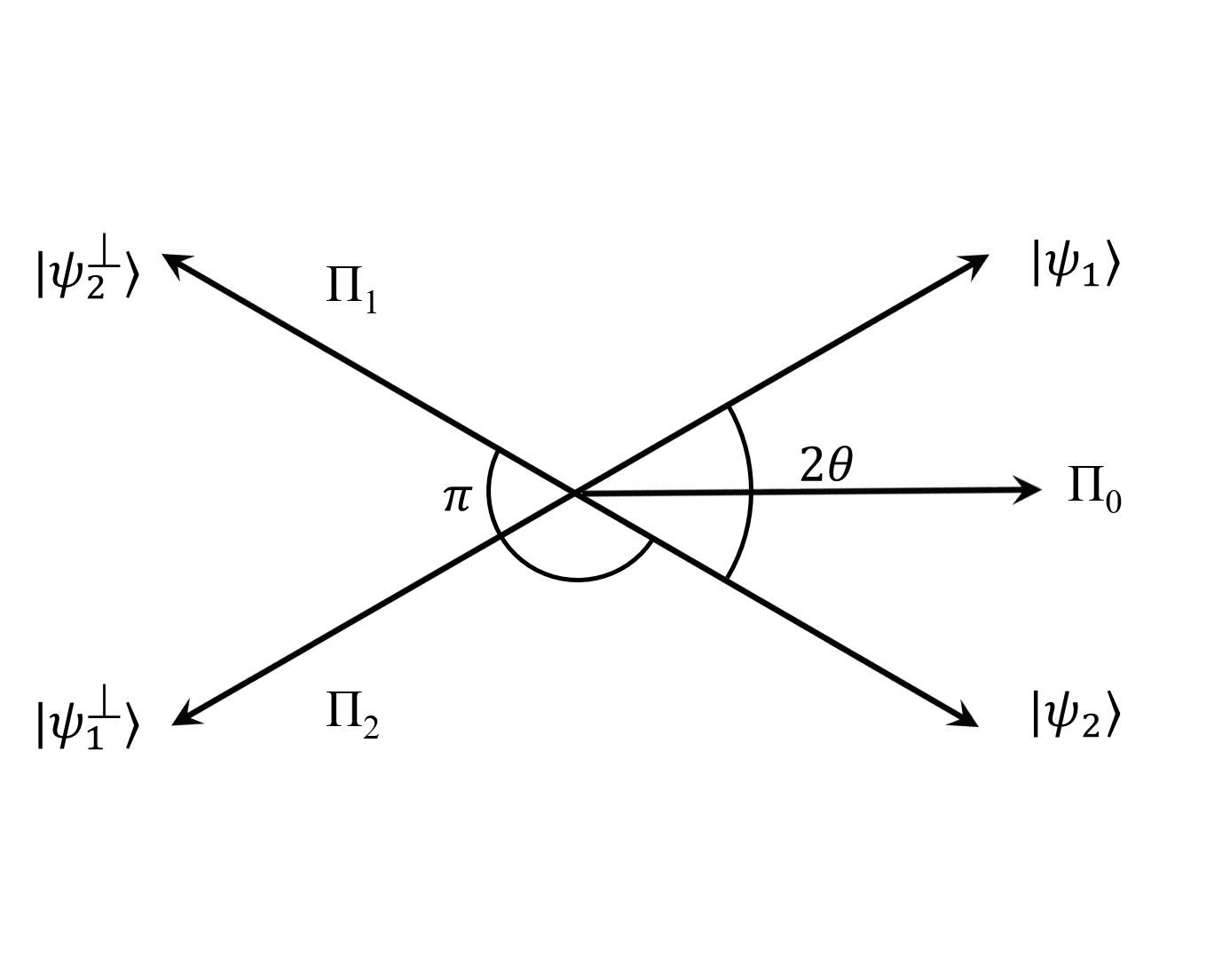}
    \caption{Bloch Sphere representation of the POVM basis. When trying to discriminate between two states $\ket{\psi_1}$ and $\ket{\psi_2}$, the directions of the $\Pi_0, \Pi_1, \Pi_2$ show the detector configuration for inconclusive, correct and incorrect detection. }
    \label{fig:POVM_basis}
\end{figure}

When trying to discriminate among non-orthogonal states, POVMs can be formulated for implementing \textbf{unambiguous state discrimination}(USD). In this approach, each outcome either perfectly recognizes a state or is inconclusive, meaning no incorrect outcome is possible, although success becomes probabilistic. The design of an optimal POVM aims to maximize the probability of successfully and unambiguously discriminating between these states. Foundational works by Helstrom, Ivanovic, Chefles, Dieks, and Peres have contributed greatly to the understanding and application of state discrimination methods, including those incorporating POVMs.

The method of unambiguous state discrimination, where an optimal POVM is applied for accurate discrimination between two quantum states, was developed through the contributions of Ivanovic, Dieks and Peres, thus earning this method the name of IDP approach to state discrimination. The POVM elements are designed as:
\begin{align*}
    \Pi_1 &= c_1 \ket{\psi_2^\perp}\bra{\psi_2^\perp}  \\
    \Pi_2 &= c_1 \ket{\psi_1^\perp}\bra{\psi_1^\perp}  \\
    \Pi_0 &= I - \Pi_1 - \Pi_2
\end{align*}
To maximize the success probability under the constraint $\Pi_0 \geq 0$ we set:
\[
c_1 = c_2 = \frac{1}{1 + \abs{\braket{\psi_1}{\psi_2}}}
\]
In this method, the probability of correct detection($P_s$) is characterized by 
\begin{equation}
    P_s = \frac{1}{2}\left(\bra{\psi_1}\Pi_1\ket{\psi_1}+\bra{\psi_2}\Pi_2\ket{\psi_2}\right)
\end{equation}
and the probability of inconclusive outcome($P_q$) is given by
\begin{equation}
    P_q=\bra{\psi_i}\Pi_0\ket{\psi_i}=1-P_s=1-\bra{\psi_i}\Pi_i\ket{\psi_i}
\end{equation}
In case of discriminating between $\ket{\psi_1}=\ket{0}, \ket{\psi_2}=\ket{+}$, the probabilities are given by,
\begin{align}
    P_s &= \frac{1}{2\left(1+\abs{\braket{0}{+}}\right)} \left( \bra{0} \ket{-}\bra{-} \ket{0} + \bra{+} \ket{1}\bra{1} \ket{+} \right) \nonumber \\
    P_s &= \frac{1}{2 \left(1 + \frac{1}{\sqrt{2}} \right)} \left( \frac{1}{2} + \frac{1}{2}\right) \nonumber \\
    P_s &= 0.292893 \label{eq: b92 eta}
\end{align}
and 
\begin{align*}
    P_q = 1 - P_s = 1- 0.292893 =0.707107
\end{align*}
\subsubsection{Helstrom Approach:}
\begin{figure}[t]
    \centering
    \includegraphics[width=0.7\linewidth]{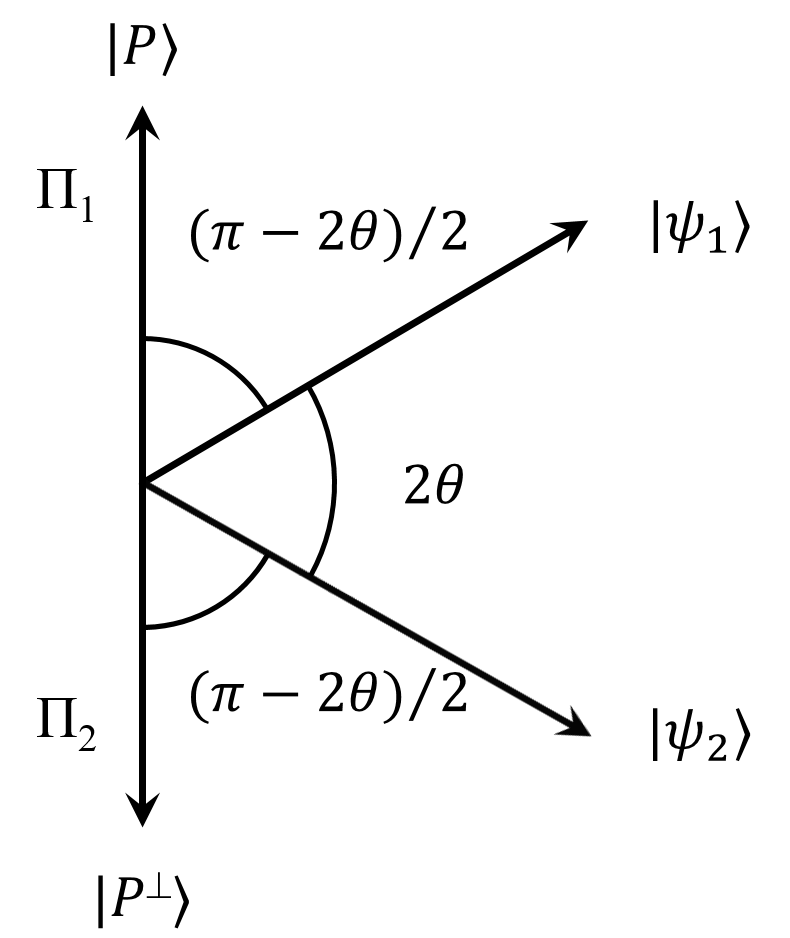}
    \caption{Bloch Sphere representation of the basis for Helstrom method of state discrimination: Detector configuration for the optimum minimum-error discrimination
    of two pure states with equal a priori probabilities. A von Neumann measurement
    with two orthogonal detectors placed symmetrically around $\ket{\psi_1}$ and $\ket{\psi_2}$
    will achieve the optimum.}
    \label{fig:Helstrom_basis}
\end{figure}

The Helstrom approach provides a strategy for minimum-error discrimination between quantum states. Unlike unambiguous discrimination, where inconclusive results are allowed to avoid errors, the Helstrom method tolerates some probability of error in exchange for maximizing the overall success rate. \newline Suppose two quantum states $\ket{\psi_1}$,$\ket{\psi_2}$ occur with prior probabilities $\eta_1,\eta_2=1-\eta_1$ respectively. The Helstrom bound gives the minimum probability of error($P_e$) when distinguishing these states as
\begin{widetext}
\begin{align*}
    P_e = \frac{1}{2}[1 - \operatorname{Tr}(\Lambda)] = \frac{1}{2}\big[ 1 - \operatorname{Tr}(\eta_2\rho_2-\eta_1\rho_1)\big] = \frac{1}{2} \big[1- ||\eta_2\rho_2-\eta_1\rho_1||\big]
\end{align*}
\end{widetext}
Fig.~\ref{fig:Helstrom_basis} shows the detector configuration for the Helstrom approach to state discrimination. In the special case where the states are pure,
\[
P_e = \frac{1}{2}(1-\sqrt{1 - 4\eta_1 \eta_2 \abs{\braket{\psi_1}{\psi_2}}^2})
\]
This expression quantifies the optimal performance achievable when no inconclusive results are allowed, and decisions must be made even under uncertainty.

The optimal measurement in this approach corresponds to a projective measurement along the eigenstates of the operator 
$\Lambda=\eta_2\rho_2-\eta_1\rho_1$
, where 
$\rho_i = \ket{\psi_i}\bra{\psi_i}.$
In case of distinguishing between pure states with equal prior probabilities($\eta_1=\eta_2=\frac{1}{2}$), the probability of error is given by 
\begin{widetext}
\begin{align*}
    P_e = \frac{1}{2}(1-\sqrt{1-\abs{\braket{\psi_1}{\psi_2}}^2}) = \frac{1}{2}(1-\sqrt{1-\cos^2\theta}) = \frac{1}{2} (1-\sin\theta)
\end{align*}
\end{widetext}
Thus, the probability of correct detection($P_s$) of state is given by, 
\[
P_s = 1-P_e = 1-(\frac{1}{2}(1-\sin\theta))=\frac{1}{2}(1+\sin\theta)
\]
Applying these formulas, one can easily estimate the probabilities in discriminating between $\ket{0}$ and $\ket{+}$. The inner angle between the states , $\theta= \arccos{\braket{0}{+}} = \arccos{\frac{1}{\sqrt{2}}}= \pi/4$. Hence, the probabilities of correct and incorrect/error outcomes are given by $P_s=\frac{1}{2}(1+\sin\pi/4)=\frac{1}{2}(1+\frac{1}{\sqrt{2}})=0.8536$ and $P_e=1-0.8536=.1464$. Note that the probability of inconclusive results, $P_q=0$ in MED.
The Helstrom approach is crucial in quantum information theory, particularly when practical applications demand deterministic outcomes—even at the cost of potential errors.
\subsubsection{FRIO Approach:}
The FRIO (Fixed Rate of Inconclusive Outcome) approach offers a hybrid strategy between minimum-error and unambiguous state discrimination. It is especially useful in scenarios where a fixed probability of inconclusive outcomes is tolerated, and the goal is to minimize the error rate within that constraint.

Unlike the Helstrom approach (which allows no inconclusive outcomes) and the USD approach (which allows only inconclusive or correct outcomes with no errors), the FRIO strategy optimally balances both error and inconclusiveness based on a predefined threshold.

Let $Q$ denote the fixed rate of inconclusive outcomes, i.e., the probability that the measurement returns an inconclusive result. The aim is to design a POVM $\{\Pi_0,\Pi_1,\Pi2\}$ where $\Pi_0$ corresponds to the inconclusive outcome, $\Pi_1$ and $\Pi_2$ correspond to conclusive identification of states $\psi_1$ and $\psi_2$, respectively.

The constraint is: $\operatorname{Tr}(\rho \Pi_0)=Q$,
and the goal is to minimize the total probability of incorrect conclusive decisions, i.e., to minimize the error probability 
$P_e$ given that the measurement is conclusive.

This leads to an optimization problem, often solved using semidefinite programming or analytical bounds depending on the geometry of the states involved. The FRIO strategy provides an interpolating continuum between the Helstrom and USD strategies:
\begin{itemize}
    \item When $Q=0$, the FRIO method reduces to Helstrom’s minimum-error discrimination.
    \item When $P_e =0$, it recovers USD’s unambiguous discrimination.
\end{itemize}
The FRIO approach is particularly useful when the user or device can tolerate some level of inconclusiveness in exchange for better reliability in conclusive identification.

\section{Generalized method of state discrimination (GSD)}
\begin{figure}[t]
    \centering
    \includegraphics[width=1\linewidth]{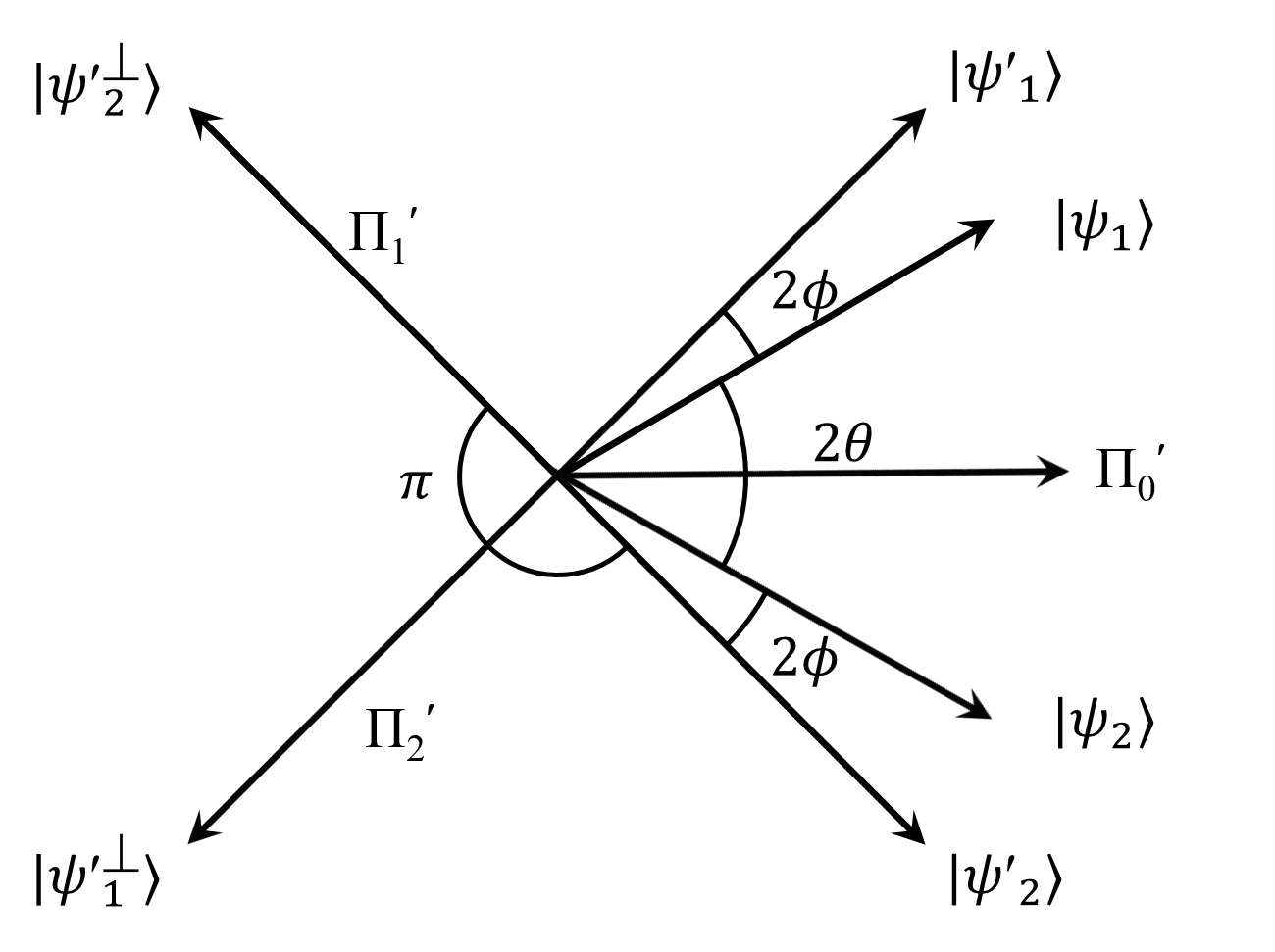}
    \caption{Bloch Sphere representation of the generalized approach to state discrimination: Here, $\ket{\psi_1},\ket{\psi_2}$ are two quantum states with inner angle $\theta$. By tilting these quantum states in an angle $\phi$, the primed quantum states i.e. $\ket{\psi_1^\prime},\ket{\psi_2^\prime}$ are found. Again, by taking the orthogonal states of the primed states the new basis states $\ket{\psi_1^{\prime \perp}},\ket{\psi_2^{\prime \perp}}$ are constructed which will be used in the formulation of the POVM elements $\Pi_1^\prime, \Pi_2^\prime$ respectively. $\Pi_0^\prime$ represents the basis for inconclusive outcomes. Note on the Bloch sphere, the angle between state vectors corresponds to twice the angle used in their quantum state representation. }
    \label{fig:General_Approach_basis}
\end{figure}
\subsection{Mathematical Formulation}
Let us consider the single qubit pure equiprobable states $\ket{\psi_1}$ and $\ket{\psi_2}$ with their inner product defined as $|\braket{\psi_1}{\psi_2}|=\cos \theta$. In USD method, the orthogonal states of the $\ket{\psi_i}$ i.e. $\ket{\psi_i^\perp}$ are taken as the basis of measurement. We modify this basis by tilting all the orthogonal states in an angle $\phi$ such that the inner angle between the newly found basis states becomes $(\theta + 2\phi)$. These new bases of measurement can be constructed by following these steps:
\begin{itemize}
    \item Tilt the $\ket{\psi_i}$ states in $\phi$ angle in such a way that the term $|\braket{\psi_1^\prime}{\psi_2^\prime}|$ is lesser than $|\braket{\psi_1}{\psi_2}|$. The new states found by tilting may be denoted as $\ket{\psi_i^\prime}$.
    \item Find out the orthogonal states of the $\ket{\psi_i^\prime}$ states. These orthogonal states $\ket{\psi_i^{\prime \perp}}$ are the proposed basis of measurement.
\end{itemize}

Fig.~\ref{fig:General_Approach_basis} shows the new tilted bases of measurement proposed in this method. Now, let us design a POVM on this $\ket{\psi_i^{\prime \perp}}$ basis. The POVM elements may be defined as follows:
\begin{equation}\label{POVM2}
\begin{aligned}
    \Pi_1^\prime &= \frac{1}{1+\abs{\braket{\psi_2^\prime}{\psi_1^\prime}}}\ket{\psi_2^{\prime \perp}}\bra{\psi_2^{\prime \perp}}\\
    \Pi_2^\prime &= \frac{1}{1+\abs{\braket{\psi_1^\prime}{\psi_2^\prime}}}\ket{\psi_1^{\prime \perp}}\bra{\psi_1^{\prime \perp}}\\
    \Pi_0^\prime &= \frac{2 \abs{\braket{\psi_2^\prime}{\psi_1^\prime}}}{1+\abs{\braket{\psi_2^\prime}{\psi_1^\prime}}}\ket{\gamma^{\prime}}\bra{\gamma^{\prime}}\\
    \gamma^{\prime} &= \frac{1}{\sqrt{2(1+\abs{\braket{\psi_2^\prime}{\psi_1^\prime}})}}(\ket{\psi_1^{\prime }}+ e^{iarg(\braket{\psi_2^\prime}{\psi_1^\prime})} \ket{\psi_2^{\prime }})
\end{aligned}
\end{equation}
Note that this formulation works only when all the POVM elements are of rank 1. After the measurement, we find three types of results:
\begin{itemize}
    \item Correct detection of state $\ket{\psi_i}$
    \item Incorrect detection of state $\ket{\psi_i}$
    \item Inconclusive outcome
\end{itemize}
This implies that if Alice sends the state $\ket{\psi_1}$ (or $\ket{\psi_2}$) then we have the following probabilities:
\begin{enumerate}
    \item Probability of detecting $\ket{\psi_1}$ as $\ket{\psi_1}$(or $\ket{\psi_2}$ as $\ket{\psi_2}$), $P_s = \left(\bra{\psi_1}\Pi_1^\prime \ket{\psi_1}+\bra{\psi_2}\Pi_2^\prime \ket{\psi_2}\right)/2$
    \item Probability of detecting $\ket{\psi_1}$ as $\ket{\psi_2}$(or $\ket{\psi_1}$ as $\ket{\psi_2}$), $P_e = \left(\bra{\psi_1}\Pi_2^\prime \ket{\psi_1}+\bra{\psi_2}\Pi_1^\prime \ket{\psi_2}\right)/2$
    \item Probability of inconclusive detection, \newline$P_q = \left(\bra{\psi_1}\Pi_0^\prime \ket{\psi_1}+\bra{\psi_2}\Pi_0^\prime \ket{\psi_2}\right)/2$
\end{enumerate}
Now let us find simpler expressions for the probabilities in terms of the inner angle($\theta$) and the tilting angle($\phi$):
\textit{Probability of correct detection or success probability, }
\begin{align}\label{correct_probability}
      P_s &= \frac{\sin^2\left( \theta + \phi\right)}{1 + \lvert\cos(\theta + 2\phi)\rvert}
    \end{align}
\textit{Probability of incorrect detection or error probability,}

\begin{align}\label{incorrect_probability}
        P_e &= \frac{\sin^2\phi}{1 + \lvert\cos(\theta + 2\phi)\rvert}
\end{align}
\textit{Probability of inconclusive outcome,}
\begin{widetext}
\begin{equation}\label{inconclusive_probability}
    \begin{aligned}
        P_q = \frac{\cos(\theta+2\phi) \big[ \cos^2\phi 
+ \cos^2(\theta + \phi) + 2 \cos\phi \cos(\theta+\phi) \cos\alpha
\big]}{[1+\abs{\cos(\theta+2\phi)}]^2} 
    \end{aligned}
\end{equation}
\end{widetext}
The detailed derivation of the simplified expressions of probabilities as shown in Eqs.~\eqref{correct_probability},\eqref{incorrect_probability}, \eqref{inconclusive_probability} can be found in Appendix~\ref{app: detailed_calculation}. 
\subsection{GSD to distinguish between $\ket{0}$ and $\ket{+}$}
\begin{figure}[t]
    \centering
    \includegraphics[width=1\linewidth]{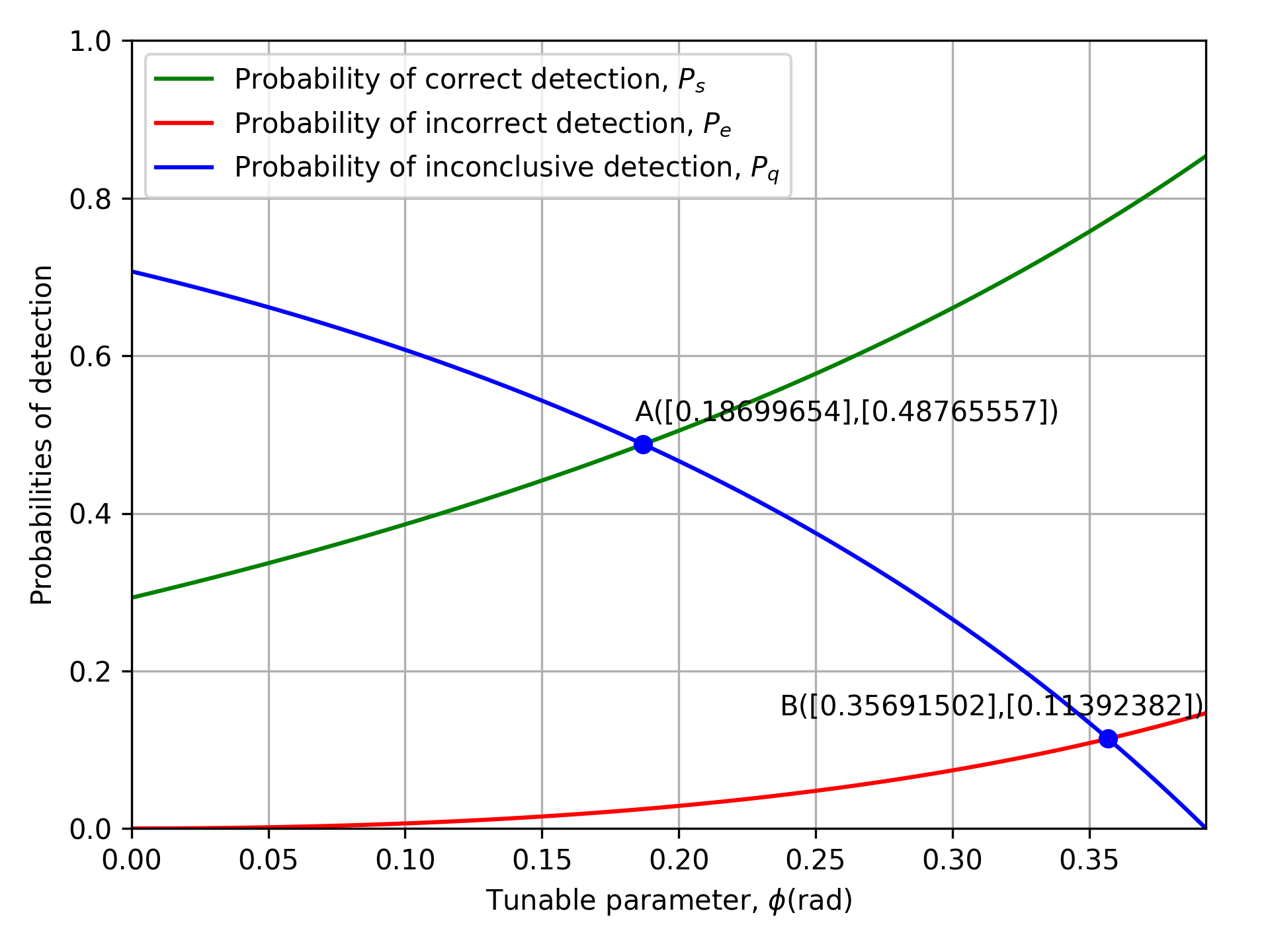}
    \caption{Probabilities vs. Tilting Angle ($\phi$): Correct detection probability ($P_s$) [Green], incorrect discrimination probability ($P_e$) [Red], and probability of inconclusive result ($P_q$) [Blue] vary smoothly and continuously in the discrimination range [USD, MED].}
    \label{fig:Prob_vs_phi}
\end{figure}

Considering $\ket{\psi_1}=\ket{0}$ and $\ket{\psi_2}=\ket{+}=\frac{1}{\sqrt{2}}\big(\ket{0}+\ket{1}\big)$, we can write :
\[\braket{\psi_1}{\psi_2} = \braket{0}{+} = \frac{1}{\sqrt{2}}\]
So, \[\alpha = \arccos(0) = \frac{\pi}{2}\]
and \[ \theta = \arccos{\frac{1}{\sqrt{2}}}= \frac{\pi}{4}\]
Now, from the Eqs.~\eqref{correct_probability},\eqref{incorrect_probability},\eqref{inconclusive_probability} we derive the following expressions of the probabilities in terms of only tilting angle $\phi$ :
\begin{equation}
    P_s = \frac{\sin^2\left(\frac{\pi}{4} + \phi\right)}{1 + \cos(\frac{\pi}{4} + 2\phi)}
\end{equation}
\begin{equation}
     P_e = \frac{\sin^2 \phi}{1 + \cos(\frac{\pi}{4} + 2\phi)}
\end{equation}
\begin{widetext}
\begin{align}
P_q &= \frac{\cos\left(\frac{\pi}{4}+2\phi\right) \left[ \cos^2\phi 
+ \cos^2\left(\frac{\pi}{4} + \phi\right) + 2 \cos\phi \cos\left(\frac{\pi}{4}+\phi\right) \cos\frac{\pi}{2}
\right]}{\left[1+\cos\left(\frac{\pi}{4}+2\phi\right)\right]^2} \nonumber \\
P_q &= \frac{\cos\left(\frac{\pi}{4}+2\phi\right) \left[\cos\phi+\cos\left(\frac{\pi}{4}+\phi\right)\right]^2}{\left[1+\cos\left(\frac{\pi}{4}+2\phi\right)\right]^2} 
\end{align}
\end{widetext}
We, then, find out the values of these probabilities with respect to a range of values of $\phi$ from $\phi_{USD}=0~\text{rad}$ up to $\phi_{MED}=\frac{\pi}{4}-\frac{\theta}{2} = \frac{\pi}{4}-\frac{\pi}{8}=\frac{\pi}{8}$ and plot a (Probabilities VS $\phi$) graph as shown in Fig.~\ref{fig:Prob_vs_phi}.

\subsection{Special Cases of GSD}
\begin{figure}[t]
    \centering
    \includegraphics[width=1\linewidth]{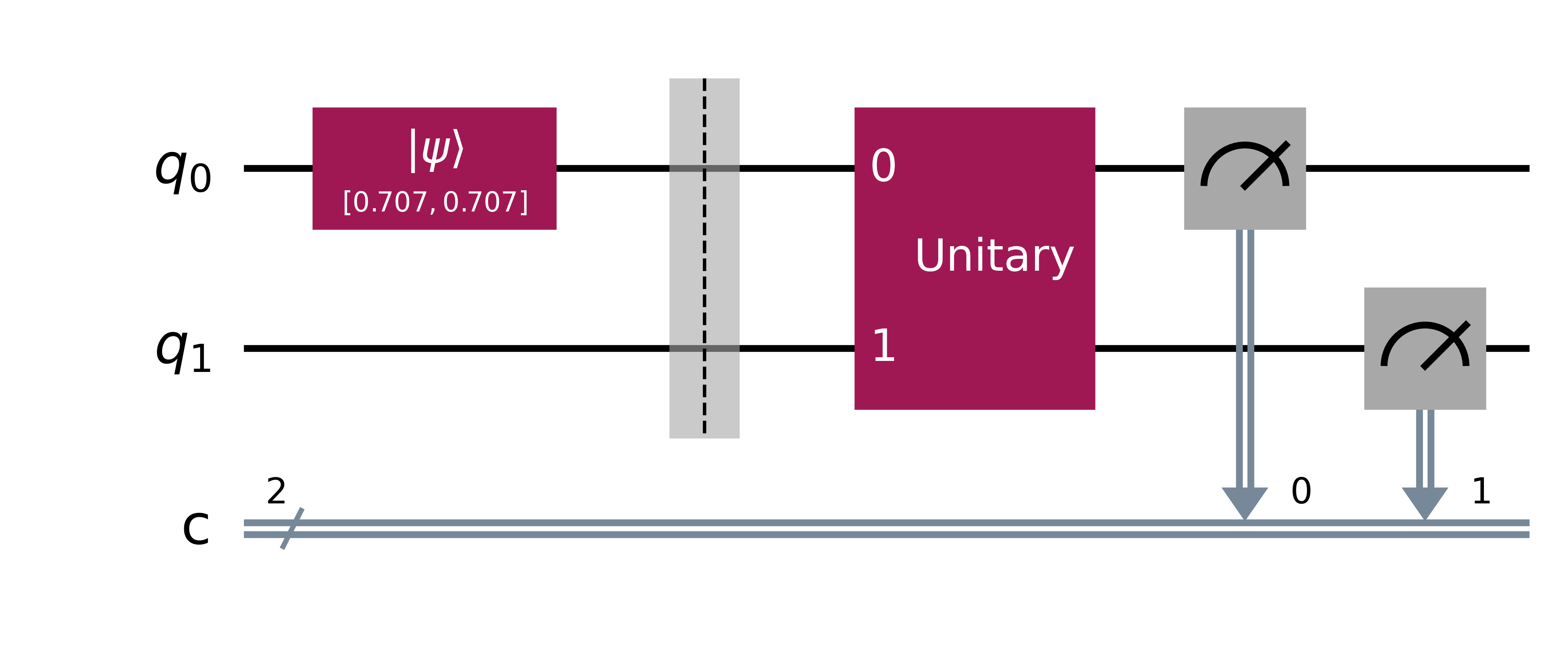}
    \caption{Quantum circuit in Qiskit for implementing generalised state discrimination (GSD). The state to be sent, in this case $\ket{\psi_2}=\ket{+}$, is prepared or initialized in qubit $q_0$ and the unitary operation $U$ and the measurements on the qubits $q_0, q_1$(an ancilla qubit) represent the measurement in the GSD proposed basis.}
    \label{fig:quantum_circuit}
\end{figure}

From Fig.~\ref{fig:Prob_vs_phi}, we find two intersection points, say, $A \equiv (0.186997, 0.487656),B \equiv (0.356915, 0.113924)$ in the region between the USD limit($\phi=0$) and the Helstrom limit($\phi=\phi_H$).
These points are of great significance since they represent the symmetry between the probabilities. \newline
\newline
\textbf{Symmetric Point A:}
\begin{itemize}
    \item This point illustrates the symmetry between the probability of correct detection and inconclusive result.
    \item  It can be interpreted as a 50/50 measurement success rate, but with very low error. It reflects a trade-off between confident guesses and abstention.
    \item From now on, we will refer to this point as the ``\textbf{Confidence Threshold Point(CTP)}'' and the corresponding tilting angle as $\phi_{CTP}$.
\end{itemize}
\textbf{Symmetric Point B:}
\begin{itemize}
    \item This point shows another form of symmetry although between incorrect and inconclusive probabilities.
    \item Neither error nor inconclusive result is prioritized - it reflects a neutral, evenly risk-balanced measurement.
    \item For further reference, we choose to call this point the ``\textbf{Equal-Risk Point(ERP)}'' and the tilting angle as $\phi_{ERP}$ 
\end{itemize}
The values of $\phi_{CTP}$ and $\phi_{ERP}$ for any two signal states $\ket{\psi_1},\ket{\psi_2}$ can be numerically obtained by following their respective condition of symmetry i.e. $P_s = P_q$ for CTP and $P_e = P_q$ for ERP. Thus, we find numerically:
\[\phi_{CTP} = 0.186997,\quad\phi_{ERP} = 0.356915\]
From Eqs.~\eqref{correct_probability},\eqref{incorrect_probability} we calculate the following probabilities at ERP:
\[
    P_s = 0.772152, \quad P_e = P_q = 0.113924.
\]
We verify the numerically found probabilities by simulating in Qiskit's AerSimulator. Fig.~\ref{fig:quantum_circuit} shows the simulation setup i.e., the quantum circuit implemented in Qiskit.
By running this quantum circuit in the AerSimulator for $10^{8}$ counts, we find similar results as the numerical ones as can be seen from Table~\ref{tab:combined_results}.

Then, we can calculate the accuracy($\chi$)[Probability of correct detection among the conclusive outcomes], and efficiency($\zeta$)[Probability of conclusive detection] as follows:
\begin{equation}
    \begin{aligned}
        \chi &= \frac{P_s \times 100}{1 - P_q} = \frac{0.772152 \times 100}{1 - 0.113924} \approx 87.14\%, \\
        \zeta &= (1 - P_q) \times 100 = (1 - 0.113924) \times 100 \approx 88.61\%.
    \end{aligned}
\end{equation}
Similarly, we can find out the accuracy and efficiency of measurements at the points USD, MED and CTP by estimating the probabilities. 

By comparing the accuracy and efficiency in the USD and Helstrom approach which are similarly calculated, we can verify our claim that GSD allows us to reach higher accuracy than MED and higher efficiency than USD, depending on the tunable parameter(tilting angle) $\phi$. The Table~\ref{tab:combined_results} shows the calculated data for the three methods and aligns with our claim.
\begin{table*}[t]
\caption{\label{tab:combined_results} Simulation results for $10^8$ shots and the corresponding accuracy ($\chi$) and efficiency ($\zeta$) across different state discrimination strategies. The exact counts correspond to correct (00), incorrect (01), and inconclusive (10) events. The GSD method is evaluated at the Confidence Threshold Point (CTP, $P_s=P_q$) and Equal Risk Point (ERP, $P_e=P_q$).}
\begin{ruledtabular}
\begin{tabular}{lccccc}
\textbf{Method} & \textbf{Correct (00)} & \textbf{Incorrect (01)} & \textbf{Inconclusive (10)} & \textbf{Accuracy} $\bm{\chi}$ \textbf{(\%)} & \textbf{Efficiency} $\bm{\zeta}$ \textbf{(\%)} \\
\colrule
Helstrom (MED) & 85,353,894 & 14,646,106 & 0 & 85.38 & 100.00 \\
IDP (USD) & 29,286,909 & 0 & 70,713,091 & 100.00 & 29.29 \\
GSD at CTP & 48,769,498 & 2,464,955 & 48,765,547 & 95.18 & 51.23 \\
GSD at ERP & 77,223,025 & 11,388,707 & 11,388,268 & 87.14 & 88.61 \\
\end{tabular}
\end{ruledtabular}
\end{table*}

\subsection{Application of GSD in QKD}
The generalized state discrimination (GSD) framework introduced in this work enables a tunable trade-off between conclusive correct detections, incorrect identifications, and inconclusive outcomes. In this section, we apply this framework to quantum key distribution by reformulating the well-known B92 protocol using a GSD-based measurement strategy.

Standard B92 relies on unambiguous state discrimination (USD) to minimize QBER(Quantum Bit Error Rate). However, the high overlap between the two signal states (typically $\ket{0}$ and $\ket{+}$) results in a large fraction of inconclusive outcomes, often exceeding 70\%, which drastically limits the protocol’s throughput. While this ideal zero-error rate is theoretically attractive, it is not always optimal in realistic implementations with lossy channels, imperfect detectors, and timing jitter.

To address this, we propose a tunable variant of B92, which we refer to as phiQKD. In phiQKD, the USD measurement is replaced with a family of hybrid POVMs parameterized by a tilt angle 
$\phi$, which controls the deviation from the standard unambiguous discrimination basis. As $\phi$ increases from 0 (USD) toward the Helstrom optimal measurement, the probability of inconclusive outcomes decreases, but at the cost of a small increase in error probability. This creates a continuous spectrum of operating points: from ultra-conservative(Higher security) to aggressive(Higher efficiency), allowing the protocol to dynamically adjust to channel noise or hardware limitations.

The performance of phiQKD is evaluated using the Devetak–Winter bound for key rate:
\[
R(\phi) \geq \eta(\phi)\left[H(X|E)(\phi)-H(X|Y)(\phi)\right]
\]
where 
$
\eta(\phi)=P_s+P_e
$
is the sifting efficiency, 
$Q(\phi) = \frac{P_e}{1 - P_q}$ is the quantum bit error rate (QBER), and $H(X|Y)$ and $H(X|E)$ represent the conditional entropies of Alice's bits given Bob’s and Eve’s knowledge, respectively. The term $H(X|Y)$ will be taken as,
\( H_{max}(X|Y) \approx H(Q) = -Q \log_2(Q)-(1-Q) \log_2(1-Q) \)
according to Shannon entropy.

We bound Eve's information using the entropic uncertainty relation \cite{renner2008security}, which yields:

\[
H(X|E) \geq log_2(\frac{1}{c})-H(X|Y)
\]
where $c=\abs{\braket{\psi_1}{\psi_2}}^2$.
\subsubsection{Analysis of Key Rates}
We now estimate the secure key rate across various values of $\phi$.
First of all, we try the numerically found values of $\phi_{ERP}$ and $\phi_{CTP}$ to find the asymptotic key rate of phiQKD.
\paragraph{Asymptotic Key Rate at ERP: ($\phi = \phi_{ERP}$)}
\begin{itemize}
    \item \(P_e = P_q = 0.113924\) and \(P_s = 0.772152\)
    \item \(\eta = 0.886076\), \(QBER, Q = 0.128571\)
    \item \(H_{max}(X|Y)=H(Q)\approx 0.553506\) and \(H_{min}(X|E) \geq 0.446494\)
    \item \(R_{\infty}(\phi_{ERP}) = -0.094521\)
    \item Key rate is negative, implying that no secure key can be generated at this point. This is mainly due to the high QBER despite achieving significant sifting efficiency.
\end{itemize}
\paragraph{Asymptotic Key Rate at CTP: ($\phi = \phi_{CTP}$)}
\begin{itemize}
    \item \(P_s = P_q = 0.487656\) and \(P_e = 0.024689\)
    \item \(\eta = 0.512344\), \(QBER, Q = 0.048188\)
    \item \(H_{max}(X|Y)=H(Q)\approx 0.278649\) and \(H_{min}(X|E) \geq 0.721351\)
    \item \(R_{\infty}(\phi_{CTP}) = 0.226816\)
    \item Key rate is reasonably good with high sifting efficiency. However, the key rate does not exceed that of the standard B92 protocol.
\end{itemize}
This motivates us to find a range of values for the asymptotic key rate with respect to different values of $\phi$ to ascertain the optimum value of $\phi$ at which configuration the highest key rate is achievable.
\begin{figure}[t]
    \centering
    \includegraphics[width=1\linewidth]{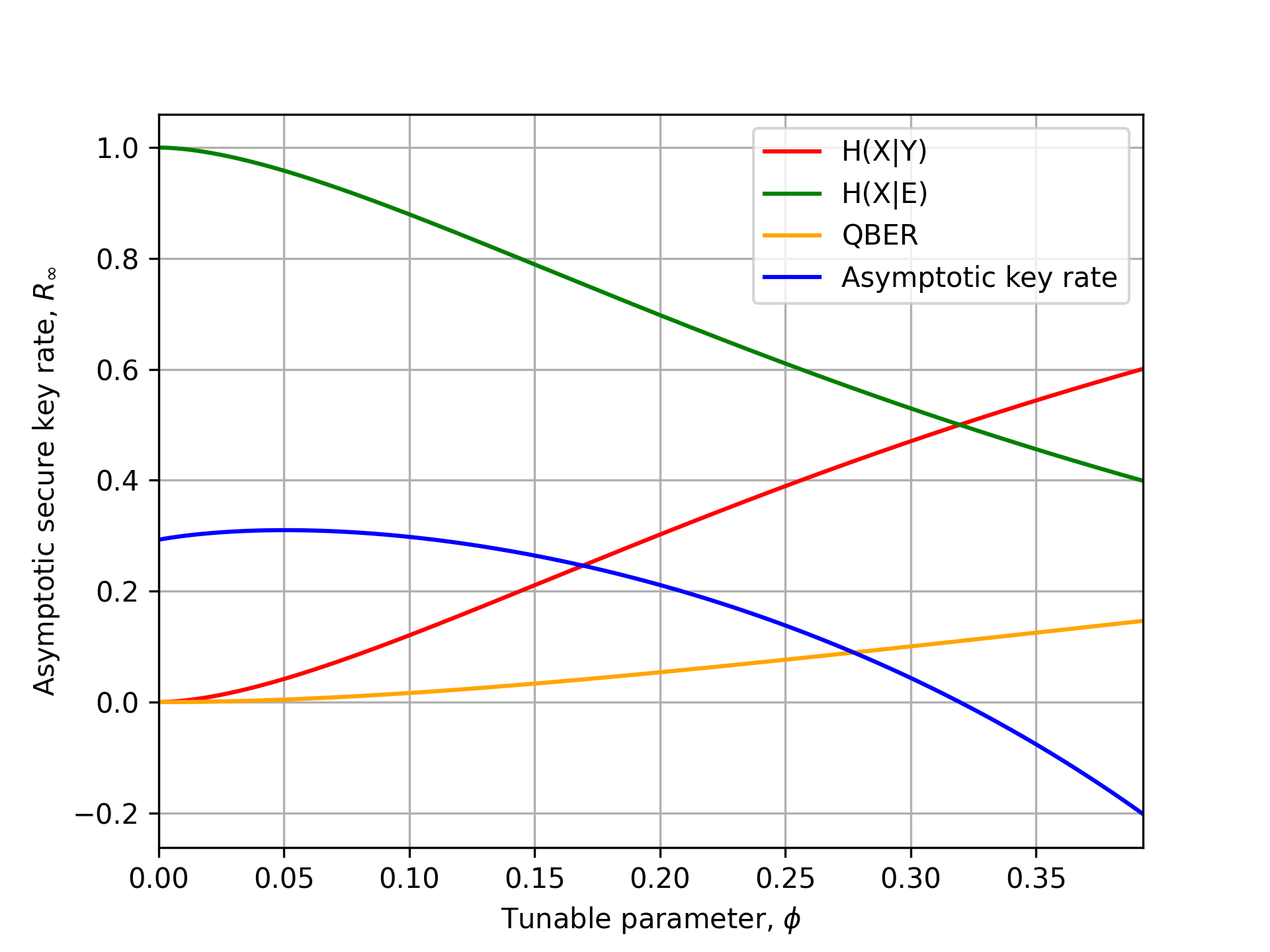}
    \caption{Plot showing the estimated asymptotic key rate $R_{\infty}$ (Blue), QBER $Q$ (Orange), $H(X|E)$ (Green), and $H(X|Y)$ (Red) for different values of $\phi$, with the signal states $\ket{0}$ and $\ket{+}$ fixed (i.e., fixed $\theta = \frac{\pi}{4}$).}
    \label{fig:keyrate-vs-phi}
\end{figure}
From Fig.~\ref{fig:keyrate-vs-phi}, we find a peak between the USD limit and the MED limit. This peak represents the highest achievable asymptotic key rate for the fixed signal states $\ket{0}$ and $\ket{+}$. Numerically solving, we find $\phi_{OPT}^{\infty} = 0.050389$. In the following section, we consider $\phi=\phi_{OPT}^{\infty}$ and find out the highest asymptotic key rate possible when the overlapping angle between the signal states is $\theta = \frac{\pi}{4}$. 
\section{The \NoCaseChange{phiQKD} Protocol}
Here, we provide a step-by-step execution plan to implement our proposed phiQKD protocol. First, Alice (the sender) prepares the signal states $\ket{\psi_j}$ based on a randomly selected bit $i \in \{0,1\}$ from a bit string where $j=i+1$. Then she sends the prepared state to our receiver Bob. Note that the choice of the two signal states is well known beforehand. We define the signal states as such: \[\ket{\psi_1}=\ket{0}, ~\ket{\psi_2}=\cos{\theta}\ket{0}+\sin{\theta}\ket{1}\]
where $\theta=\abs{\braket{\psi_1}{\psi_2}}$ is the overlap angle between the signal states and is known by both parties. Now, Bob decides the values of the following parameters:
\begin{itemize}
    \item Tilting angle, $\phi \in \{\phi_{USD},\phi_{Helstrom}\}$ where $\phi_{USD}=0~rad,~\phi_{Helstrom}=\frac{\pi}{2}-\frac{\theta}{2}$ 
    \item Probabilities of correct $(P_{s})$, incorrect $(P_{e})$ and inconclusive $(P_{q})$ outcomes for each value of parameter $\phi$
    \item Sifting efficiency, $\eta=P_{s}+P_{e}=1-P_{q}$
    \item Quantum bit error rate (QBER), $Q=\frac{P_{e}}{\eta}$
    \item Number of total signals $(N)$ and sample bits for parameter estimation $(n)$
    \item Hoeffding correction parameter, $\delta \leq \sqrt{ \frac{1}{2 \cdot n} \ln\left( \frac{2}{\epsilon_{pe}} \right) }$
    \item Worst probable QBER, $Q_{worst}\leq Q+\delta$
    \item The security parameters: failure probability in parameter estimation $(\epsilon_{pe})$, secrecy failure probability $(\epsilon_{sec})$ and correctness failure probability $(\epsilon_{cor})$
    \item Error correction efficiency $(f)$
    \item Estimated composable secure key rate 
    \begin{widetext}
    \begin{equation}\nonumber
        R_{secure} \leq \frac{1}{N}\left[(n_{sifted}-n) \cdot \left(\log_{2}\left(\frac{1}{c}\right) - H(Q_{\text{worst}})\right) - leak_{EC}- \log_2\left(\frac{2}{\epsilon_{sec}^{2}\epsilon_{cor}}\right)\right]
    \end{equation}
    \end{widetext}
    where $n_{sifted}=N\cdot \eta$, $c=\abs{\braket{\psi_1}{\psi_2}}^{2}$ and $leak_{EC}=(n_{sifted}-n)\cdot f\cdot H(Q_{worst})$
    \item The value of $\phi=\phi_{opt}$ for which the highest possible estimated secure key is found.
\end{itemize}
Then Bob uses $\phi_{opt}$ to design the GSD POVM elements and measures the state sent to him. Upon discarding the inconclusive outcome associated bits, the sample bits used in parameter estimation, the leakage bits while doing error correction, a secure key is finally generated.
\subsection{Derivation of Key Rates for phiQKD}
In this section, we provide the explicit calculation steps for the highest asymptotic, finite-key, and composable secure key rates for the phiQKD protocol. We assume the signal states \( |0\rangle \) and \( |+\rangle \), with overlapping term \( c = 0.5 \). A tilting angle \( \phi=\phi_{OPT}^{\infty}=0.050389~rad \) is considered to find out the highest asymptotic key rate as follows:

\begin{table}[b]
\centering
\caption{Success, error, and inconclusive probabilities with efficiency $\eta$ for $\phi = \phi_{OPT}^{\infty}$.}
\label{tab:probabilities_infinite}
\begin{ruledtabular}
\begin{tabular}{cccc}
$P_{s}$ & $P_{e}$ & $P_{q}$ & $\eta$ \\
\hline
0.337118 & 0.001554 & 0.661328 & 0.338657 \\
\end{tabular}
\end{ruledtabular}
\end{table}
\subsubsection{ Asymptotic Key Rate}
Using $Q = \frac{0.001554}{0.337118+0.001554} \approx 0.004588$:
\begin{itemize}
    \item \( H_{max}(X|Y) \approx H(Q) = -Q \log_2(Q)-(1-Q) \log_2(1-Q) \approx 0.042228 \)
    \item From Entropic Uncertainty Relation, \( H_{min}(X|E) \geq  0.957772\)
\end{itemize}
This gives us the asymptotic key rate,
\begin{align}
R_{\infty}(\phi) &\geq \eta\left[H(X|E)(\phi) - H(X|Y)(\phi)\right] = 0.310055
\end{align}

\subsubsection{ Finite-Key Rate (Hoeffding Corrected)}
From Hoeffding's inequality for binary state \cite{hoeffding1963probability}:
\begin{equation}
\operatorname{Pr}\big[\lvert Q-\hat{Q} \rvert \geq\delta\big] \leq 2 e^{-2n\delta^{2}}
\end{equation}
Let, $2e^{-2n\delta^2} = \epsilon_{pe}$, where $\epsilon_{pe} $ is failure probability in parameter(Q) estimation. We consider $\epsilon_{pe} = 10^{-10}$ as standard for this particular formulation. Here, n is the number of samples used for parameter estimation. Now solving for $\delta$ we get,
\begin{equation}
\delta \leq \sqrt{ \frac{1}{2 \cdot n} \ln\left( \frac{2}{\epsilon_{pe}} \right) }
\end{equation}
\begin{align}
Q_{\text{worst}} \leq Q + \delta
\end{align}
Here, we consider sampling without replacement for the Hoeffding bound. Hence, finite key rate,
\begin{equation}\nonumber
R_{\text{finite}} \geq \left(\eta -\frac{n}{N}\right) \cdot \left[H(X|E) - H(Q_{worst})\right]
\end{equation}
where $N$ is the total number of signal qubits sent by Alice. The subtraction of n is due to the discarding of the sample qubits used in parameter estimation.
Further analysis reveals that the optimal value of parameter $\phi$ for which the highest finite key rate is found is actually $\phi_{OPT}^{finite}=0.083261~rad$.

\begin{table}[htbp]
\centering
\caption{Success, error, inconclusive probabilities and sifting efficiency for $\phi=\phi_{OPT}^{finite}$ considering sample size $N$.}
\label{tab:probabilities_finite}
\begin{ruledtabular}
\begin{tabular}{ccccc}

$P_{s}$ & $P_{e}$ & $P_{q}$ & $\eta$ & $N$ \\
\hline
0.368881 & 0.004377 & 0.626742 & 0.373258 & $10^{6}$ \\
\end{tabular}
\end{ruledtabular}
\end{table}
\vspace{0.5em}
We find, $Q = \frac{0.004377}{0.368881+0.004377} \approx 0.011726 $. In our evaluation, we take $n = 10^{5}$, which is roughly 30\% of sifted qubits.

\[
\delta \leq \sqrt{ \frac{1}{2 \cdot 10^5} \ln\left( \frac{2}{10^{-10}} \right) } = 0.010890
\]

The maximum possible quantum bit error rate or the worst-case QBER, 
\begin{align}
Q_{\text{worst}} \leq Q + \delta = 0.011726 + 0.010890 = 0.022616 \nonumber
\end{align}
The finite secure key rate,
\begin{align}
R_{\text{finite}} &\geq (\eta - \frac{n}{N}) \cdot (H(X|E) - H(Q_{worst})) = 0.188063
\end{align}
\subsubsection{ Composable Secure Key Rate}
\begin{itemize}
    \item Sifted bits, $n_{sifed} = \eta\cdot N$ where $\eta$ is the sifting efficiency and $N$ is the number of quantum signals sent by Alice. We consider $N = 10^6$ for pedagogical reasons only.
    \item Error correction efficiency, $f=1.15$ (considered as standard for this formulation)
\end{itemize}
We set a total security parameter $\epsilon_{total}$ and split it among the failure probabilities as $\epsilon_{total} = \epsilon_{pe} + \epsilon_{cor} + \epsilon_{sec}$. In this work we use $\epsilon_{pe} = \epsilon_{cor} = \epsilon_{sec} = 10^{-10}$, i.e., $\epsilon_{total} \leq 3\times10^{-10}$. These choices are conservative and in line with finite-key QKD literature \cite{renner2008security,tomamichel2012tight}. If a more stringent $\epsilon_{total}$ is required for a particular application, the same formulae apply after replacing the $\epsilon$ values.
The final secure key length is given by,
\begin{widetext}
\begin{equation}\label{eq: secure key length}
\ell \leq (n_{sifted}-n) \cdot \left(\log_{2}\left(\frac{1}{c}\right) - H(Q_{\text{worst}})\right) - leak_{EC}- \log_2\left(\frac{2}{\epsilon_{sec}^{2}\epsilon_{cor}}\right) 
\end{equation}
\end{widetext}

We consider several parameters for the analysis, starting with the preparation quality of the source, $q = \log_{2}\left(\frac{1}{c}\right)=1$, which depends on the overlap of the signal states. We also set the failure probability $\epsilon_{cor} = 10^{-10}$ and the secrecy failure probability $\epsilon_{sec} = 10^{-10}$. The information leakage during error correction is given by $leak_{EC}=(n_{sifted}-n)\cdot f\cdot H(Q_{worst})$

To demonstrate, we assume the error correction efficiency $f=1.15$, corresponding to experimental values found for LDPC codes \cite{scarani2009security},\cite{dixon2014high} . The optimal tilting angle for which the highest secure key rate is found is $\phi_{OPT}^{secure}=0.073953~rad$. The relevant probabilities and rates are $P_{s} = 0.359635$, $P_{e} = 0.003422$, $P_{q} = 0.636946$, and $\eta = 0.363054$.\\
From these, we calculate $Q \approx 0.009426$. \\
Finally, with $\delta = 0.010890$, we can establish an upper bound for the worst-case quantum bit error rate as $Q_{\text{worst}} \leq Q + \delta = 0.009426 + 0.010890 = 0.020316$. The secure key length is then
$\ell \leq 181958.0554$, and the composable secure key rate is
$R_{\text{secure}} \leq 0.181958$.

\subsection{Comparison Between Secure key rates of B92 and phiQKD}
Now, we apply the same parameter estimation conditions for the B92 protocol: $\eta_{B92}=0.292893 $ (from subsection 3.5.2 ), QBER, $Q_{B92} = 0$, and \newline $Q_{worst} \leq Q_{B92} + \delta = 0.01089$.\\
The final secure key rate in the case of the B92 protocol becomes,
\newline $R_{B92} \leq 0.156862$
\begin{widetext}
\begin{equation}\nonumber    
\text{Improvement in secure key rate} = \frac{0.181958-0.156862}{0.156862}\times 100\% = 15.998776 \% \approx 16 \%
\end{equation}
\end{widetext}
\begin{figure*}[t]
    \centering
    \includegraphics[width=0.8\linewidth]{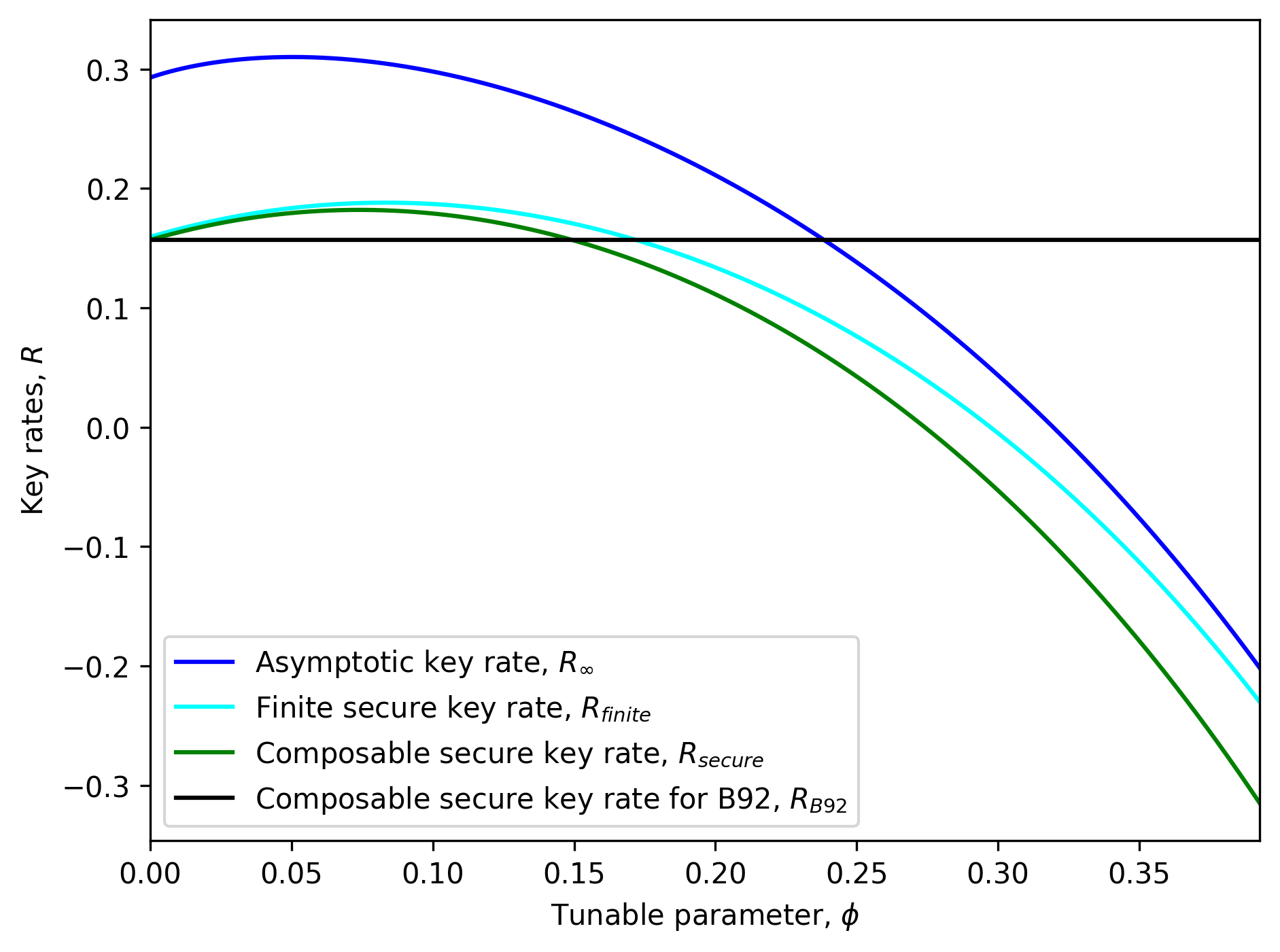}
    \caption{Expected asymptotic, finite, and composable secure key rates for phiQKD and B92 scheme as a function of the tilting angle or the tunable parameter ($\phi$). The relation that the secure key rate decreases with available qubits for the formulation holds true. For a range of values of the tilting angle, i.e., $\phi$, the composable secure key rate of the phiQKD protocol is greater than that of B92 due to higher sifting efficiency resulting from the design of the protocol.}
    \label{fig:Key_rates_VS_phi}
\end{figure*}
For the commonly studied signal state pair $\ket{0}$ and $\ket{+}$, we find that the highest composable secure key rate occurs at a tilting angle $\phi \approx 0.073953~rad$. At this point, the quantum bit error rate (QBER) remains about 2\%, and the sifting efficiency improves over B92’s 29.29\% baseline to approximately 36.3\%. While the absolute increase in secure key rate is modest (about 16\%) this gain is accompanied by significantly reduced error and greater measurement efficiency, marking a meaningful operational improvement. As can be seen from Fig.~\ref{fig:Key_rates_VS_phi}, our proposed phiQKD performs better than B92 in terms of secure key generation rate for a range of values of tunable parameter $\phi$. 
\subsubsection{The Range Parameter}
For further analysis, we define another parameter called "Range" defined as, 
\begin{align}
Range = \frac{\phi_{bound}-\phi_{USD}}{\phi_{Helstrom}-\phi_{USD}}\times 100\%
\end{align}
where, $\phi_{bound}=$~The value of $\phi$ for which the composable secure key rate of phiQKD becomes equal to the key rate of B92 and continues to fall.
Notice that, $\phi_{USD}=0$ holds for any value of overlapping angle $\theta$. And hence,
\[
Range(\theta) = \frac{\phi_{bound}}{\phi_{Helstrom}}\times 100\%
\]
For $\theta = \frac{\pi}{4}~rad$ we have $\phi_{bound}=0.149123~rad$
    and $\phi_{Hestrom}=\frac{\pi}{8}~rad$, so 
\[
Range(\theta=\frac{\pi}{4}) = \frac{0.149123}{\frac{\pi}{8}}\times 100\%=37.97\%
\]
This implies that for approximately 38\% of the range between the USD and Helstrom limit, our proposed phiQKD protocol shows better key generation rate than the B92 protocol. However, this analysis is only considered for the case when $\theta=\frac{\pi}{4}$. Further analyses are required to truly establish a benchmark of the proposed protocol.
\subsubsection{Sensitivity to Overlap Angle}

We further analyze the protocol’s performance across a range of signal overlap angles $\theta \in \{\pi/2\}$.
\paragraph{Range Analysis}
According to the definition of the signal states $\ket{\psi_1},\ket{\psi_2}$ we consider a range of values of $\theta \in \{0,\frac{\pi}{2}\}$ and find out the ``Range'' for each values of the overlapping angle $\theta$. 
\begin{figure}[t]
    \centering
    \includegraphics[width=1\linewidth]{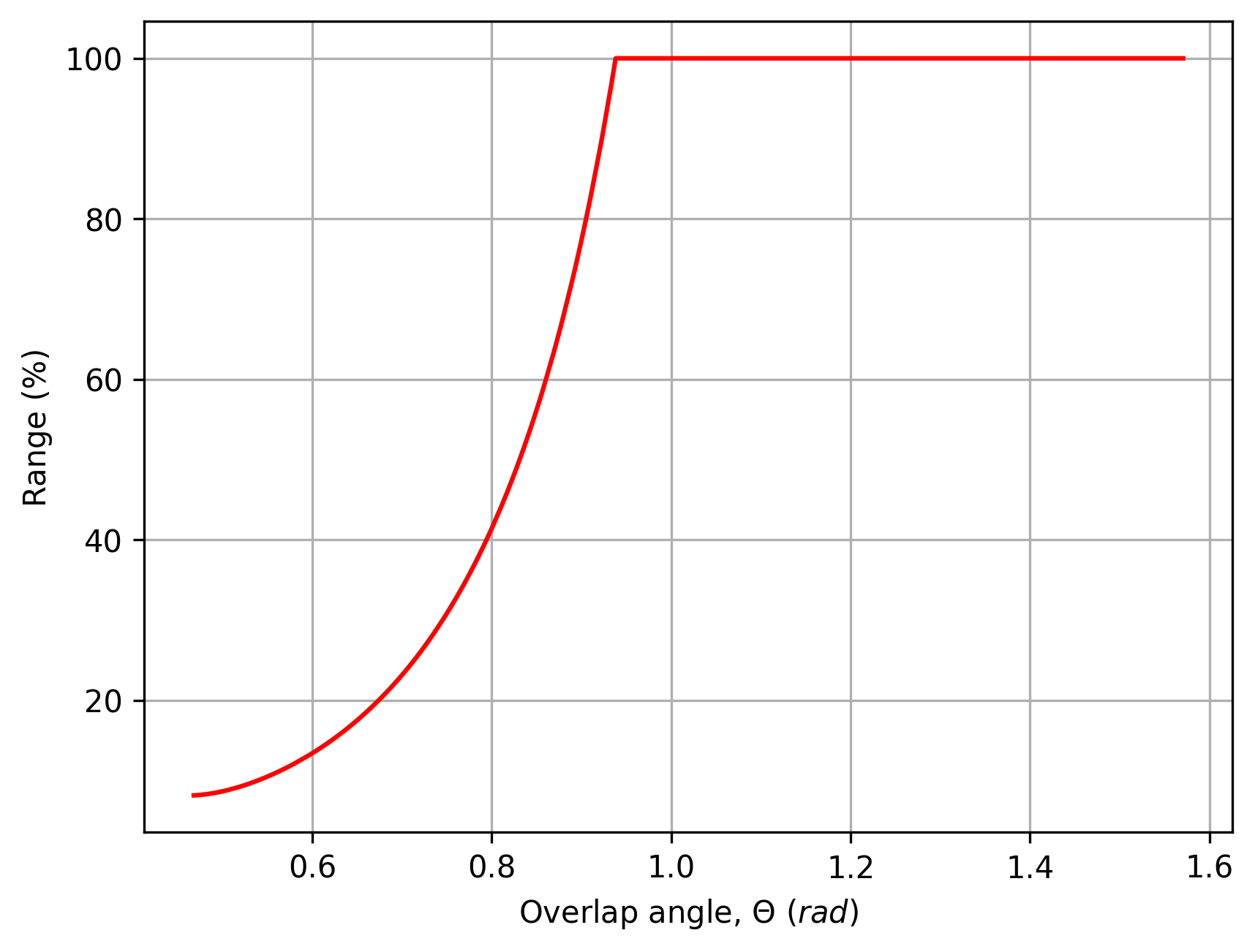}
    \caption{Range i.e. the percentage range of values of the tiling angle for which the composable secure key rate in phiQKD is greater than that in the B92 scheme as a function of the Overlap angle, $\theta$.}
    \label{fig:percentage}
\end{figure}
As can be seen from Fig.~\ref{fig:percentage}, the Range increased almost exponentially until it reached 100\% at $\theta = 0.938015~rad$ and stayed the same until $\theta = \frac{\pi}{2}$. This implies that for $\theta \in \{0.938015,\frac{\pi}{2}\}$ phiQKD shows improved key rates than the B92 protocol for any value of $\phi \in \{\phi_{USD},\phi_{Helstrom}\}$. 
\paragraph{Optimality of Tilting Angle Depending on Overlap}

\begin{figure}
    \centering
    \includegraphics[width=1\linewidth]{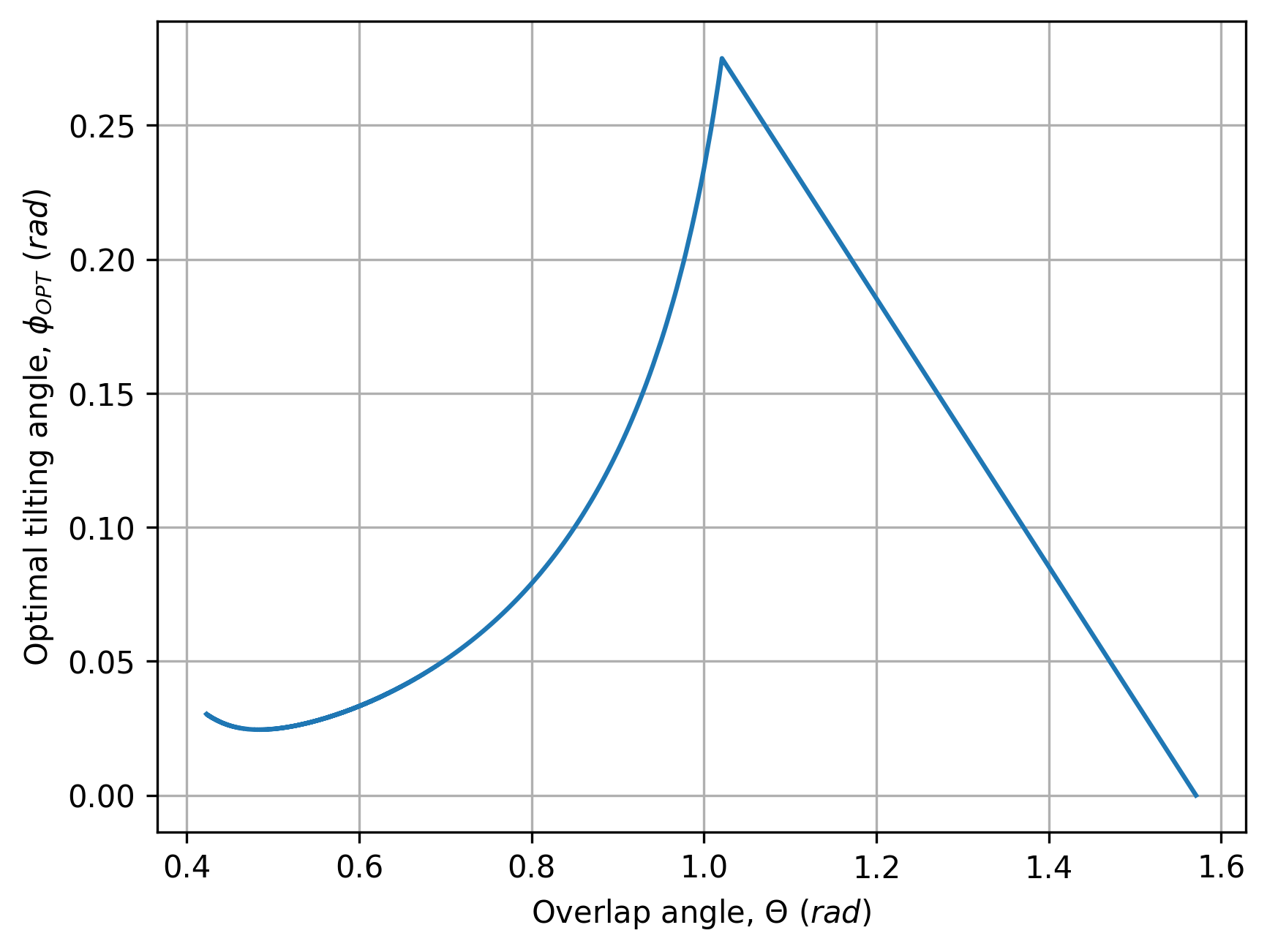}
    \caption{Optimal tilting angle, $\phi_{OPT}$ (The tilting angle for which the highest composable secure key rate is found in phiQKD) as a function of the overlap angle $\Theta$.}
    \label{fig:phi_opt}
\end{figure}

We also found out the value of $\phi_{OPT}$ for any value of $\theta \in \{0,\frac{\pi}{2}\}$ for which a positive secure key rate exists. In Fig.~\ref{fig:phi_opt}, we see that the value of $\phi_{OPT} \in \{0,0.274995\}$. This is a crucial discovery, since it shows that by tweaking the USD basis by a very small angle (about 16 degrees), we can achieve a much-improved key rate in the modified B92 protocol, i.e., the phiQKD protocol. At the boundary point i.e. for $\theta=\frac{\pi}{2}$ we find that $\phi_{OPT}=0~rad$. This also agrees with our formulation since at $\theta = \frac{\pi}{2}$ the projective measurement is sufficient to perfectly distinguish between two states.
\subsubsection{Highest Secure Key Rate Comparison Between phiQKD and B92}
For these values of $\phi_{OPT}$, we now find out the highest secure key rates possible for our phiQKD protocol and B92 protocol.  
\begin{figure}[t]
    \centering
    \includegraphics[width=1\linewidth]{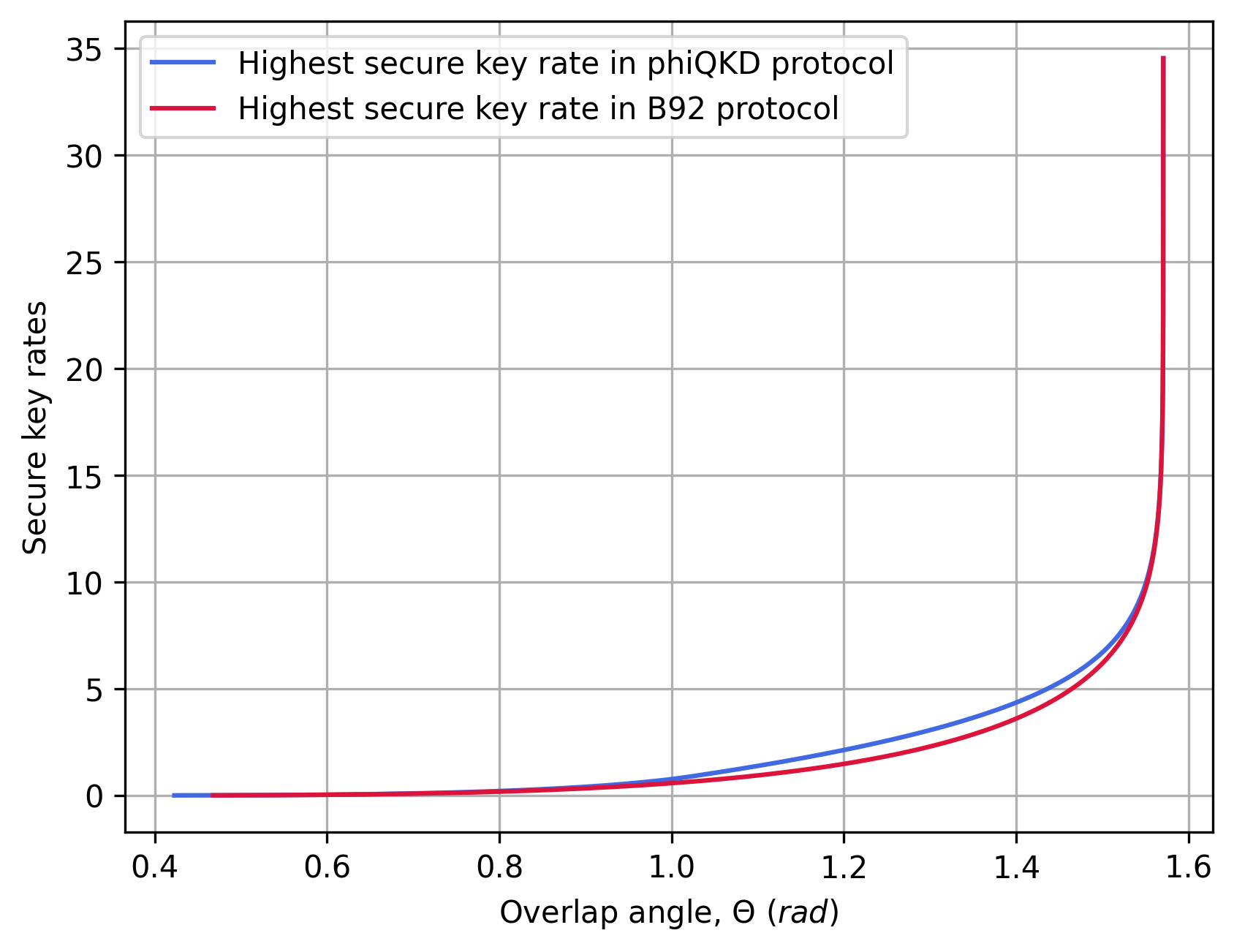}
    \caption{Highest positive composable secure key rates for phiQKD and B92 protocols with increasing values of the \mbox{Overlap} angle, $\theta$. For some lower values of $\theta$, phiQKD generates a secure key, but B92 scheme does not.}
    \label{fig:skr_vs_theta}
\end{figure}
Fig.~\ref{fig:skr_vs_theta} shows that for a good range of values of $\theta$, the secure key rate of phiQKD is greater than B92 protocol. There are even some lower values of $\theta$, where secure key generation is possible in phiQKD, although not possible with B92 scheme.

Now, we find out the difference in secure key rates between the two schemes in the following way,
\[\text{Difference} = R_{secure}^{phiQKD}-R_{secure}^{B92}\]
where $R_{secure}^{phiQKD}$ and $R_{secure}^{B92}$ represent the highest possible secure key rate in our scheme and the B92 protocol respectively.
\begin{figure}[b]
    \centering
    \includegraphics[width=1\linewidth]{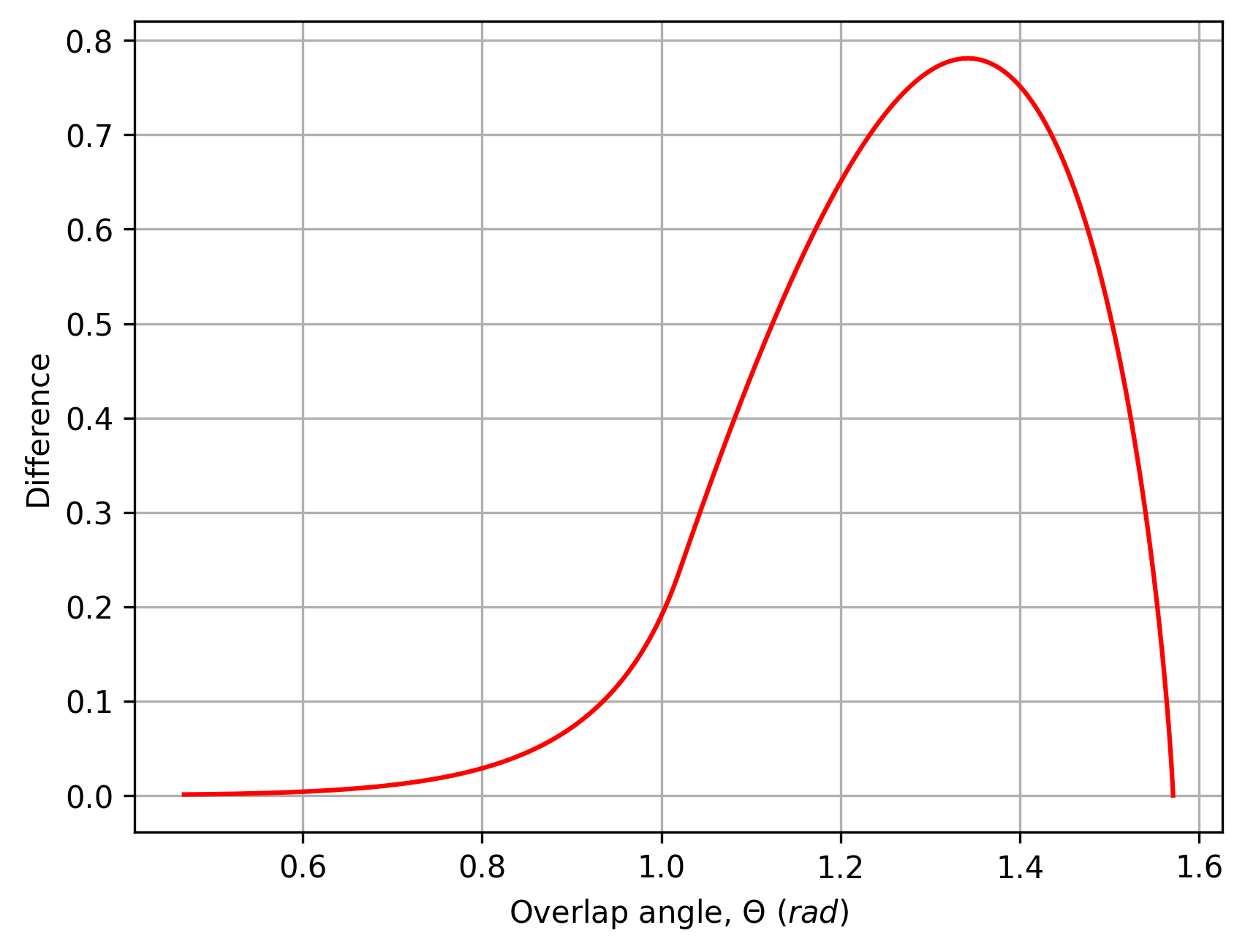}
    \caption{Difference between the highest possible composable secure key rates between the phiQKD and B92 protocols with respect to a range of values of overlap angle}
    \label{fig:difference_vs_theta}
\end{figure}
From Fig.~\ref{fig:difference_vs_theta}, We can find the highest difference in key generation rate of about 0.781095 at $\theta = 1.341750~rad$. 
\begin{figure}[t]
    \centering
    \includegraphics[width=1\linewidth]{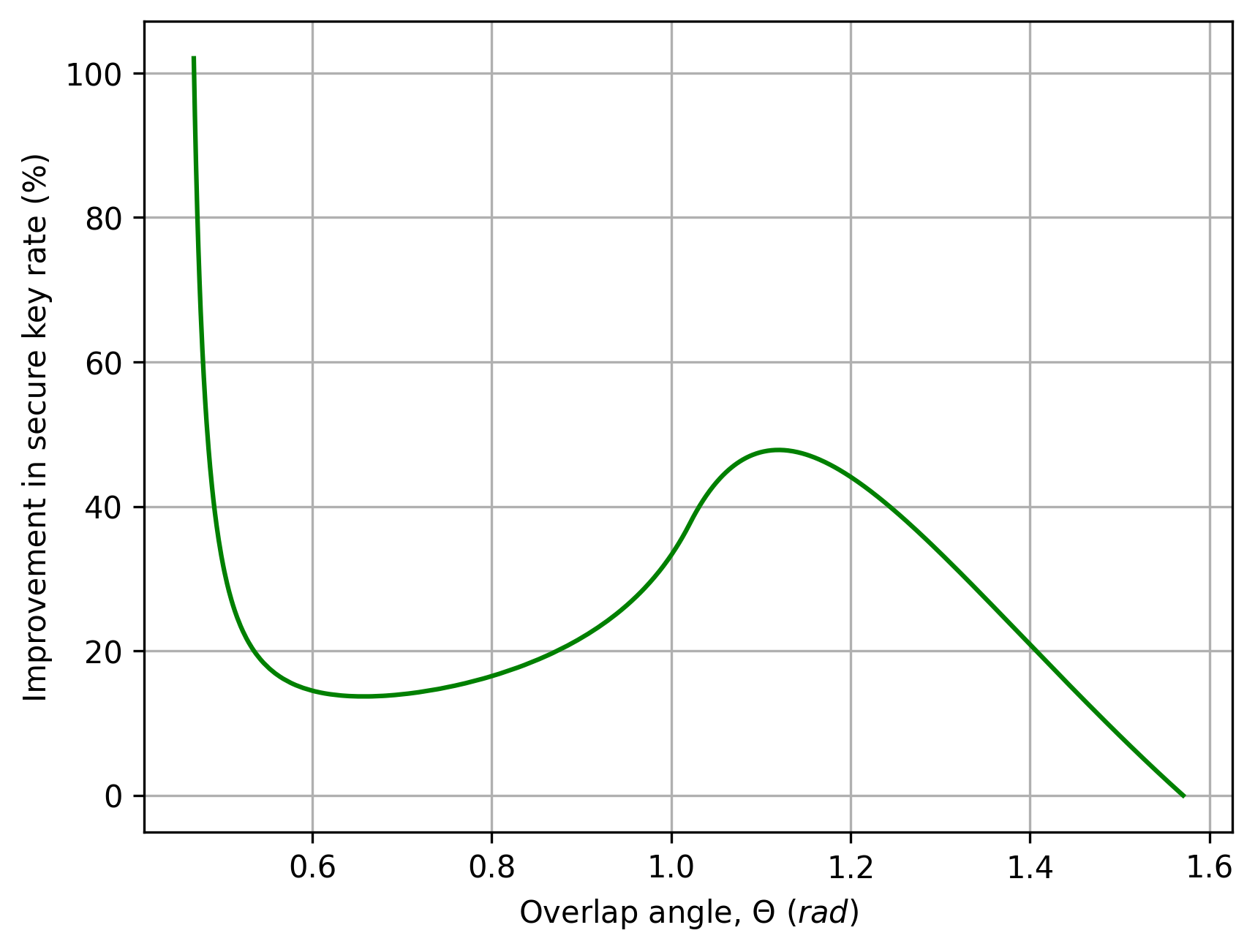}
    \caption{Improvement (\%) in the highest composable secure key rate in the phiQKD protocol compared to the B92 scheme with Overlap angle, $\Theta$. To show a meaningful graph we have considered those values of overlap angle for which the highest secure key rate in B92 scheme is not less than 0.001.}
    \label{fig:improvement_vs_theta}
\end{figure}
To get a comparative sense of improvement, we find out the improvement (\%) in secure key rate in phiQKD with respect to B92 for those values of $\theta=\{0,\frac{\pi}{2}\}$ for which there is a positive key in the B92 scheme. Fig.~\ref{fig:improvement_vs_theta} illustrates that although the improvement is quite high in lower values of $\theta$, this is due to the value of key rates being very small. We see some credible improvement in the upper ranges of values of overlap angle and find the highest improvement of about 47.82\% at $\theta=1.119617~rad$. 


In summary, our simulations show that phiQKD:
\begin{itemize}
    \item Outperforms standard B92 by 16\% in standard configurations, and for a certain range of tilting angles between the USD and MED limits, always demonstrates better key rates.
    \item Provides up to 47.82\% improvement in a specific configuration of non-orthogonal signal states.
    \item Shows positive secure key rates in conditions where fixed-basis protocols, like B92, fail.
    \item Offers a mathematically predictable and geometrically intuitive framework for adaptability in practical protocol implementation.
\end{itemize}

\section{Conclusion}
We have proposed a generalized state discrimination (GSD) strategy and applied it to the B92 protocol to create phiQKD, a tunable protocol whose key parameter is the measurement basis tilt $\phi$. Unlike standard B92, which fixes its measurement as an ideal USD, phiQKD introduces a continuous family of POVMs that interpolate between unambiguous and minimum-error discrimination. This tunability allows the protocol to adapt to channel noise and implementation imperfections, yielding secure key rates even in conditions where traditional B92 would become impractical.

\noindent While the maximal key rate exceeds B92’s theoretical limit only slightly, the core contribution of this work is not the numerical improvement, but the introduction of a physically realizable and dynamically controllable measurement framework. This framework not only enables optimized performance in existing two-state protocols, but also lays the foundation for adaptive, noise-aware QKD systems, and for future generalizations beyond qubits. In this way, phiQKD demonstrates how rethinking quantum measurements as tunable resources, rather than fixed primitives, can lead to more robust and flexible quantum communication protocols.

\section{Discussion and Outlook}
The framework of generalized state discrimination (GSD) developed in this work solidifies the perspective on how quantum measurements can be leveraged as tunable resources in protocol design, rather than static procedures. By applying this to quantum key distribution through the phiQKD protocol, we have demonstrated that modest improvements in secure key rate can be accompanied by significant gains in robustness, adaptability, and practical usability.

Unlike traditional QKD protocols, which operate at fixed measurement points (e.g., USD in B92 or projective bases in BB84), phiQKD introduces a continuously adjustable measurement configuration, defined by a single geometric parameter $\phi$. This allows the protocol to respond dynamically to channel noise, photon loss, and device imperfections, enabling more reliable operation in realistic environments where ideal assumptions often break down. Beyond QKD, the GSD approach may find applications in quantum sensing, quantum metrology, and quantum hypothesis testing, wherever binary quantum state discrimination is a limiting task. In these contexts, the ability to navigate a continuous space between error-free and conclusive measurements could enable new trade-offs between precision and confidence.

While we only show an example focusing on a specific qubit pair ($\ket{0},\ket{+}$), the GSD formalism is general and can be extended to arbitrary pairs of pure states with non-zero overlap. Future work could explore designing an experimental setup to realize the GSD measurement and making the protocol adaptive, where the measurement basis is tuned in real-time based on observed error rates or channel statistics. Optimal tilting strategies for other state configurations (e.g., B92 with biased states, or multi-state generalizations) can also be studied along with machine learning integration, to automatically learn optimal tilting parameters in variable environments.

In conclusion, this work encourages a shift in how we think about measurement in quantum information. Instead of thinking of it as a fixed operation, we can think of it as a flexible, designable, and controllable process that can be shaped to suit the problem at hand. Proven by the example of the phiQKD protocol, this change in perspective leads to far more efficient and robust quantum protocols against channel imperfections and noise. Dealing with such constraints is crucial as we move towards building the quantum internet. This research contributes to this endeavor by focusing on the realization of adaptable and controllable quantum processes instead of static ones.

\section{AUTHOR DECLARATIONS}
\subsection{Conflict of Interest}
The authors have no conflicts to disclose.

\section{Data Availability}
The data that support the findings of this study are openly available in Zenodo at \href{https://doi.org/10.5281/zenodo.18723984}{https://doi.org/10.5281/zenodo.18723984} \cite{animesh_banik_2026_18723984}.

\clearpage
\appendix
\onecolumngrid
\section{Detailed derivation of GSD probabilities}
\label{app: detailed_calculation}
In this appendix, we provide the step-by-step algebraic derivation of the probabilities of correct, incorrect, and inconclusive detections.
\textit{Probability of correct detection or success probability, }
\begin{align}\label{app_correct_probability}
      P_s &= \left(\bra{\psi_1}\Pi_1^\prime \ket{\psi_1}+\bra{\psi_2}\Pi_2^\prime \ket{\psi_2}\right)/2\nonumber\\
      P_s &= \frac{1}{2}\left(\frac{1}{1+\abs{\braket{\psi_2^\prime}{\psi_1^\prime}}}\braket{\psi_1}{\psi_2^{\prime \perp}}\braket{\psi_2^{\prime \perp}}{\psi_1}+\frac{1}{1+\abs{\braket{\psi_1^\prime}{\psi_2^\prime}}}\braket{\psi_1}{\psi_2^{\prime \perp}}\braket{\psi_2^{\prime \perp}}{\psi_1}\right)\nonumber\\ 
      P_s &= \frac{\abs{\braket{\psi_1}{\psi_2^{\prime \perp}}}^2+\abs{\braket{\psi_2}{\psi_1^{\prime \perp}}}^2}{2\left(1+\abs{\braket{\psi_2^\prime}{\psi_1^\prime}}\right)}\nonumber\\ 
      P_s &= \frac{\cos^2\left(\frac{\pi}{2} - \theta - \phi\right)}{1 + \lvert\cos(\theta + 2\phi)\rvert} \nonumber\\
      P_s &= \frac{\sin^2\left( \theta + \phi\right)}{1 + \lvert\cos(\theta + 2\phi)\rvert}
    \end{align}
\textit{Probability of incorrect detection or error probability,}
\begin{align}\label{app_incorrect_probability}
        P_e &= \left(\bra{\psi_1}\Pi_2^\prime \ket{\psi_1}+\bra{\psi_2}\Pi_1^\prime \ket{\psi_2}\right)/2\nonumber\\
        P_e &= \frac{1}{2}\left(\frac{1}{1+\abs{\braket{\psi_1^\prime}{\psi_2^\prime}}}\braket{\psi_1}{\psi_1^{\prime \perp}}\braket{\psi_1^{\prime \perp}}{\psi_1}+\frac{1}{1+\abs{\braket{\psi_2^\prime}{\psi_1^\prime}}}\braket{\psi_2}{\psi_2^{\prime \perp}}\braket{\psi_2^{\prime \perp}}{\psi_2}\right)\nonumber\\ 
        P_e &= \frac{\abs{\braket{\psi_1}{\psi_1^{\prime \perp}}}^2+\abs{\braket{\psi_2}{\psi_2^{\prime \perp}}}^2}{2\left(1+\abs{\braket{\psi_1^\prime}{\psi_2^\prime}}\right)}\nonumber\\ 
        P_e &= \frac{\cos^2\left(\frac{\pi}{2} - \phi\right)}{1 + \lvert\cos(\theta + 2\phi)\rvert}\nonumber \\
        P_e &= \frac{\sin^2\phi}{1 + \lvert\cos(\theta + 2\phi)\rvert}
\end{align}
\textit{Probability of inconclusive outcome,}

\begin{align}\label{P_q_initial}
        P_q &= \left(\bra{\psi_1}\Pi_0^\prime \ket{\psi_1}+\bra{\psi_2}\Pi_0^\prime \ket{\psi_2}\right)/2\nonumber\\
        P_q &= \frac{1}{2}\left(\frac{2 \abs{\braket{\psi_2^\prime}{\psi_1^\prime}}}{1+\abs{\braket{\psi_2^\prime}{\psi_1^\prime}}}\braket{\psi_1}{\gamma^\prime}\braket{\gamma^\prime}{\psi_1}+\frac{2 \abs{\braket{\psi_1^\prime}{\psi_2^\prime}}}{1+\abs{\braket{\psi_1^\prime}{\psi_2^\prime}}}\braket{\psi_2}{\gamma^\prime}\braket{\gamma^\prime}{\psi_2}\right)\nonumber\\
        P_q &= \frac{1}{2}\left(\frac{2 \cos(\theta+2\phi)}{1+\cos(\theta+2\phi)}\braket{\psi_1}{\gamma^\prime}\braket{\gamma^\prime}{\psi_1}+\frac{2 \cos(\theta+2\phi)}{1+\cos(\theta+2\phi)}\braket{\psi_2}{\gamma^\prime}\braket{\gamma^\prime}{\psi_2}\right)
\end{align}
Now, 
\begin{align*}
    \braket{\psi_1}{\gamma^\prime} &= \frac{1}{\sqrt{2(1+\abs{\braket{\psi_2^\prime}{\psi_1^\prime}})}}\big(\braket{\psi_1}{\psi_1^\prime}+e^{i\arg(\braket{\psi_2^\prime}{\psi_1^\prime})}\braket{\psi_1}{\psi_2^\prime}\big)\\
\text{or, }\braket{\psi_1}{\gamma^\prime} &= \frac{1}{\sqrt{2(1+\cos(\theta+2\phi))}}\big[\cos\phi+e^{i\alpha}\cos(\theta+\phi)\big]
\end{align*}
where, \(\arg(\braket{\psi_1}{\psi_2})=\arg(a+ib) = \arctan(\frac{b}{a}) = \alpha\). Similarly,
\begin{equation} \nonumber
\braket{\gamma'}{\psi_1}
= \frac{1}{\sqrt{2(1+\cos(\theta+2\phi))}}
\left[\cos\phi + e^{-i\alpha}\cos(\theta+\phi)\right]
\end{equation}
This gives us,

\begin{align}\label{inner_gamma_1}
\braket{\psi_1}{\gamma^\prime} \braket{\gamma^\prime}{\psi_1}
&= \frac{\Big[ \cos\phi + e^{i \alpha} \cos(\theta+\phi) \Big]  \Big[ \cos\phi + e^{-i \alpha} \cos(\theta+\phi) \Big]}{2(1+\abs{\cos(\theta+2\phi)})} 
 \nonumber \\
\text{or, } \quad
\braket{\psi_1}{\gamma^\prime} \braket{\gamma^\prime}{\psi_1}
&= \frac{\cos^{2}\phi + \cos^{2}(\theta+\phi) +\cos\phi \cos(\theta+\phi) \left(e^{i \alpha} + e^{-i \alpha}\right)}{2(1+\abs{\cos(\theta+2\phi)})} \nonumber \\
\text{or, } \quad
\braket{\psi_1}{\gamma^\prime} \braket{\gamma^\prime}{\psi_1}
&= \frac{
\cos^2\phi 
+ \cos^2(\theta + \phi) 
+ 2 \cos\phi \cos(\theta+\phi) \cos\alpha}
{2(1+\abs{\cos(\theta+2\phi)})}
\end{align}

In the same manner, 

\begin{align} \label{inner_gamma_2}
    \braket{\psi_2}{\gamma^\prime} \braket{\gamma^\prime}{\psi_2}
&= \frac{
\cos^2\phi 
+ \cos^2(\theta + \phi) 
+ 2 \cos\phi \cos(\theta+\phi) \cos\alpha}
{2(1+\abs{\cos(\theta+2\phi)})}
\end{align}

Hence, we can write from Eq.~\eqref{P_q_initial},\eqref{inner_gamma_1} and \eqref{inner_gamma_2},

\begin{equation}\label{app_inconclusive_probability}
    \begin{aligned}
        P_q = \frac{\cos(\theta+2\phi) \big[ \cos^2\phi 
+ \cos^2(\theta + \phi) + 2 \cos\phi \cos(\theta+\phi) \cos\alpha
\big]}{[1+\abs{\cos(\theta+2\phi)}]^2} 
    \end{aligned}
\end{equation}

\clearpage
\twocolumngrid
\bibliography{refs}

\end{document}